\UseRawInputEncoding
\documentclass[twocolumn]{aastex631}
\usepackage{mathptmx}
\usepackage{longtable}
\usepackage{multirow}
\usepackage{listings}
\usepackage{amsmath}  
\usepackage{color,soul}
\usepackage{comment}

\makeatletter
\newcommand*{\bigcdot}{}
\DeclareRobustCommand*{\bigcdot}{%
  \mathbin{\mathpalette\bigcdot@{}}%
}
\newcommand*{\bigcdot@scalefactor}{.5}
\newcommand*{\bigcdot@widthfactor}{1.15}
\newcommand*{\bigcdot@}[2]{%
  \sbox0{$#1\vcenter{}$}
  \sbox2{$#1\cdot\m@th$}%
  \hbox to \bigcdot@widthfactor\wd2{%
    \hfil
    \raise\ht0\hbox{%
      \scalebox{\bigcdot@scalefactor}{%
        \lower\ht0\hbox{$#1\bullet\m@th$}%
      }%
    }%
    \hfil
  }%
}
\makeatother

\shorttitle{TESS Transit Timing}
\shortauthors{Ivshina \& Winn}
\graphicspath{{./}{figures/}}

\begin{document}

\title{TESS Transit Timing of Hundreds of Hot Jupiters}

\correspondingauthor{Ekaterina S.\ Ivshina}
\email{ivshina@princeton.edu}

\author[0000-0003-4457-4797]{Ekaterina S.\ Ivshina}
\affiliation{Department of Mathematics, Princeton University, Princeton, NJ 08544, USA}

\author[0000-0002-4265-047X]{Joshua N.\ Winn}
\affiliation{Department of Astrophysical Sciences, Princeton University, Princeton, NJ 08540, USA}

\submitjournal{The Astrophysical Journal Supplement Series}
\accepted{February 2, 2022}

\begin{abstract}
We provide a database of transit times and updated ephemerides for 382 planets based on data from the NASA Transiting Exoplanet Survey Satellite (TESS) and previously reported
transit times which were scraped from the literature in a semi-automated fashion. In total, our database contains 8{,}667 transit timing measurements for 382 systems. About 240 planets in the catalog are hot Jupiters (i.e. planets with mass~$>$0.3\,$M_{\rm Jup}$ and period~$<$10~days) that have been observed by TESS. The new ephemerides are useful for scheduling follow-up observations and searching for long-term period changes.  WASP-12 remains the only system for which a period change
is securely detected. We remark on other cases of interest, such as a few systems with suggestive (but not yet convincing) evidence for period changes, and the detection of a second transiting planet in the NGTS-11 system. The compilation of light curves, transit times, ephemerides, and timing residuals are made available online, along with the Python code that generated them (visit \url{https://transit-timing.github.io}).

\end{abstract}

\section{Introduction} \label{sec:intro}

Transiting hot Jupiters are the most intensively studied category of exoplanets.
Photometric and spectroscopic observations of transits
provide information about the planet's size, mass, orbit, and
atmosphere.  Because many of the follow-up observations are
time-critical, observers need the ability to predict future transit times reliably
and precisely.  This is especially important when observations need
to be scheduled with expensive and oversubscribed facilities such
as NASA's {\it James Webb Space Telescope}.
The uncertainty in transit-time predictions grows approximately
in proportion to the time elapsed since the last observation.
Thus, the easiest way to improve our ability to predict
future transits is to measure new transit times,
a process known as ``refreshing the ephemerides.''

Long-term transit timing of hot Jupiters might
also lead to the detection
of interesting physical effects, such as
planetary mass loss \citep{Valsecchi_2015,Jackson2016};
tidal orbital decay \citep[see, e.g.,][]{2013A&A...551A.108M,Patra_2020, Yee_2019}
and other forms of spin-orbit coupling \citep{Lanza_2020};
orbital precession due to
a companion planet \citep{MiraldaEscude2002},
the planet's tidal deformation \citep{WolfRagozzine2009},
general relativity \citep{Jordn2008, Pl2008};
and forces from wide-orbiting companions \citep{Bouma_2019}.
Short-timescale transit timing variations can signal the presence of nearby planets \citep{2005Sci...307.1288H,2005MNRAS.359..567A},
although these are rarely detected for hot Jupiters \citep{Steffen_2012}. 

The NASA \textit{Transiting Exoplanet Survey Satellite}
(TESS; \citealt{Ricker_2014}) offers an opportunity to measure new and precise
transit times for essentially all of the known hot Jupiters orbiting stars
brighter than about 13th magnitude.  We have taken this opportunity
to create a transit time database for 278 planets that have been
observed during the first few years of the TESS mission.
These planets are mainly hot Jupiters, though we also selected other types of
planets for which we expected the signal-to-noise ratio of
TESS observations to be high.
We also searched the literature for all previously reported transit
times to allow for improvement in the orbital ephemerides;
in this step we included an additional 104 planets that are
well-suited for TESS observations but for which TESS data are not
yet available. 

The paper is organized as follows. Section \ref{sec:sample} describes
our target selection. Section \ref{sec:tess} describes our
largely automated analysis of TESS data.
Section \ref{sec:literature} explains how
we assembled lists of transit times drawn from the literature.
Section \ref{sec:tt} describes the process we used to
refine the transit ephemerides and search for transit timing variations.
Section \ref{sec:special} discusses systems of special
interest. Section \ref{sec:summary} summarizes the results.

\section{Sample selection} \label{sec:sample}

One of the most comprehensive and carefully curated catalogs of transiting
planet properties is TEPCat, which is maintained
by John Southworth.\footnote{\url{https://www.astro.keele.ac.uk/jkt/tepcat/};
\cite{2011MNRAS.417.2166S}} For each system
in TEPCat, we estimated the signal-to-noise ratio (SNR) with which
TESS would be able to detect a single transit. The noise estimates were based
on the charts provided in the TESS Data Release Notes
showing the Combined Differential Photometric
Precision on one-hour timescales (CDPP$_1$) as
a function of apparent magnitude in the TESS bandpass.
For each system,
we looked up the TESS apparent magnitude and the corresponding
value of CDPP$_1$, and calculated
\begin{equation}
{\rm SNR} = \frac{(R_p/R_\star)^2}{{\rm CDPP}_1/\sqrt{T_{\rm hr}}},
\end{equation}
where $R_p$ and $R_\star$ are the planetary and
stellar radii, respectively, and $T_{\rm hr}$ is the total transit duration in hours.

\begin{figure*}
\begin{center}
\includegraphics[width=1\textwidth]{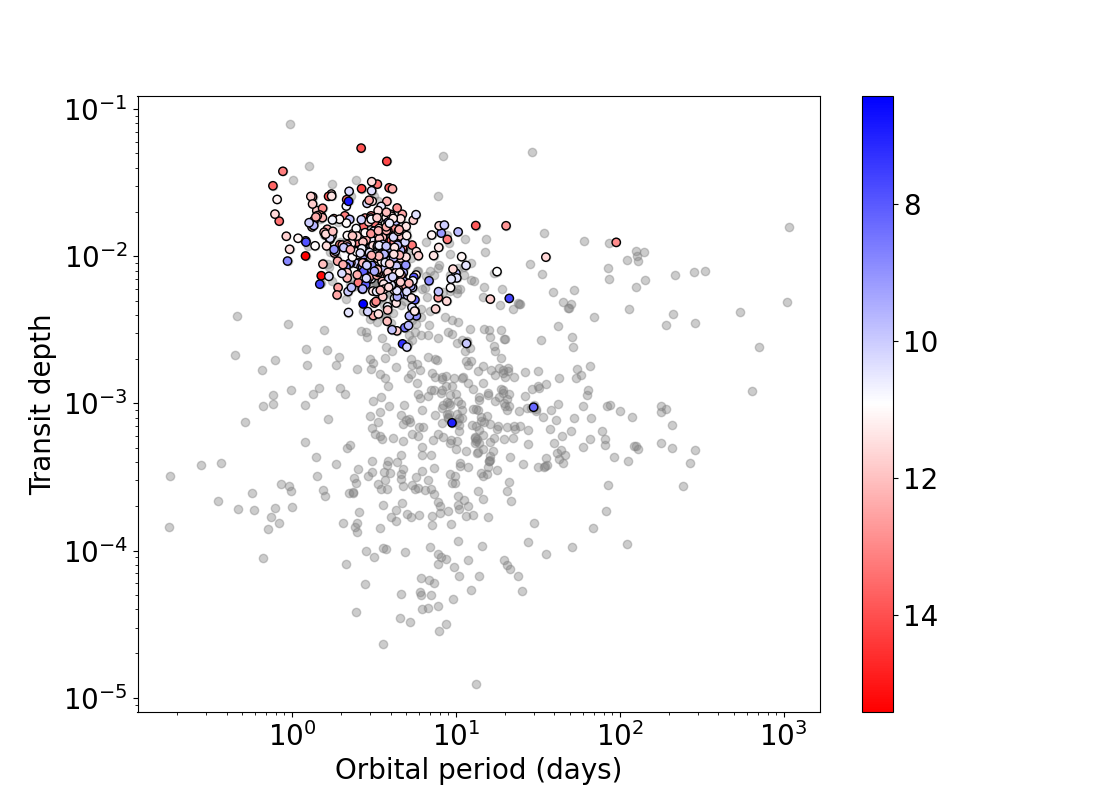}
\end{center}
\caption{Transit depth and orbital period for all of the known
transiting planets (gray), and the systems in our sample (colored points,
with the color conveying the TESS apparent magnitude).}
\label{fig:depth_vs_period}
\end{figure*}

We identified 421 systems for which SNR$~>25$.
We decided to exclude the 102 systems in this list
that involved brown dwarfs or for which no improvement was
possible because the transits were first discovered using TESS data.
We also decided to include all of the transiting planets
detected in the HAT, WASP, CoRoT, NGTS, and Qatar surveys
regardless of SNR, which led to the addition of 62 systems
with SNR~$<25$. Thus, our sample consists mainly of hot Jupiters
(mass~$>$0.3\,$M_{\rm Jup}$ and period~$<$10~days), although
it also includes some smaller and wider-orbiting planets
for which TESS data are expected to be useful.
The final list of 381 planetary systems with 382 planets is provided in
Table~\ref{tbl:target-list}.
Figure~\ref{fig:depth_vs_period} displays
the distribution of their transit depths, orbital periods, and
apparent magnitudes.
Naturally, almost all of the systems are short-period giant
planets around bright stars.

\begin{table*}
	\centering
{
\hskip-2.5cm\begin{tabular}{r|ccc}
	\hline 
System & TIC  & TESS magnitude & SNR \\ \hline
CoRoT-01 & 36352297  & 13.05 & 31.66 \\
CoRoT-02 & 391958006 & 11.67 & 111.23\\
	\hline 
\end{tabular}} 
\caption{Target List.
}
\tablecomments{Only a portion of this table is shown here to demonstrate its form and content. A machine-readable version of the full table can be accessed from the electronic version of this work.}

\label{tbl:target-list}

\end{table*}

\begin{table*}
	\centering
{
\hskip-2.5cm\begin{tabular}{r|ccc}
	\hline 
System & Source & Cadence (sec) & Sector \\ \hline
CoRoT-01 & SPOC & 120  & 33 \\
CoRoT-01 & SPOC & 120 & 6 \\
	\hline 
\end{tabular}} 
\caption{TESS Data Processed for Each Target.
}
\tablecomments{Only a portion of this table is shown here to demonstrate its form and content. A machine-readable version of the full table can be accessed from the electronic version of this work.}

\label{tbl:tess-sources}

\end{table*}

\section{TESS data analysis} \label{sec:tess}

TESS data were available for 278 of the selected systems as of
November 2021, the somewhat arbitrary cutoff date for this project. 
We wrote an automated data-reduction code to download
the TESS data and process the light curves.
We used routines from the \textsc{lightkurve} Python package \citep{2018ascl.soft12013L} to download any available light curves.
Whenever possible, we used the PDC-MAP version
of the light curves prepared by the
Science Processing Operations Center \citep{Jenkins2016}.
When these were not available, we used the light curves
from the MIT Quick-Look Pipeline \citep{qlp}.
We used light curves with 2-minute time sampling (''cadence'')
whenever possible. Otherwise, we used the data with 30-minute
cadence. The only exceptions were HATS-42 (Sector 33, 34), HATS-61 (Sector 31), HATS-70 (Sector 33, 34), WASP-182 (Sector 27), WASP-143 (Sector 35), HATS-38 (Sector 35), HATS-55 (Sector 34), HATS-64 (Sector 35), KELT-19 (Sector 33), WASP-110 (Sector 27), HATS-32 (Sector 29), WASP-160 (Sector 33), HAT-P-64 (Sector 32), for which
we used the data with a 10-minute cadence, a relatively new mode
of data collection for TESS.
In Table~\ref{tbl:tess-sources}, we specify the TESS data that we processed for each target, indicating the source and cadence of the light curves as well as the TESS sector number.
 

For each system, and for each sector of TESS data,
we identified the time intervals centered on 
predicted transit times and spanning four full transit durations.
Initially, the predicted transit times were not always accurate and the intervals were not
well centered on the transit mid-points,
because of the accumulated uncertainty in the published
ephemerides.
After performing
the first iteration of our entire analysis and refreshing the ephemerides,
we repeated everything with the updated predictions for the transit
times, to ensure that the intervals in this step were centered on the transits.

\begin{figure*}

\begin{center}
\includegraphics[width=1\textwidth]{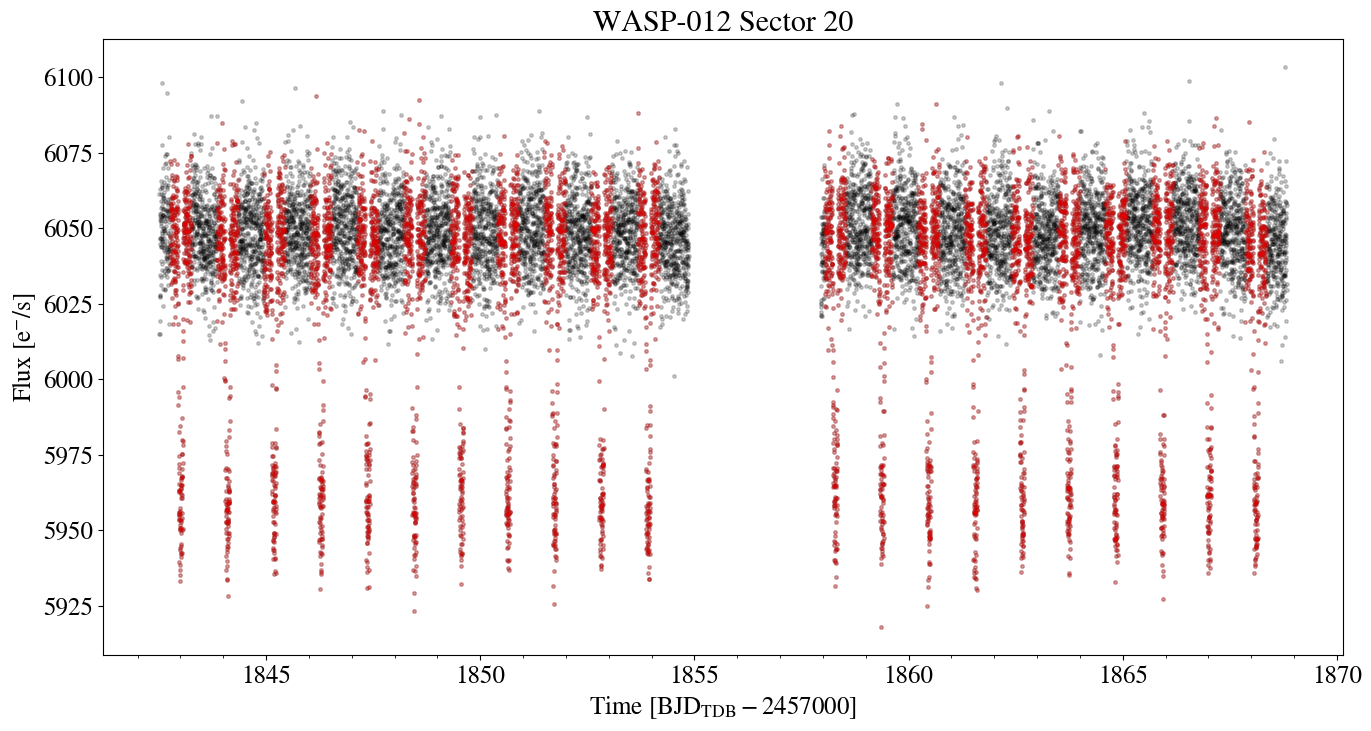}
\end{center}
\caption{TESS light curve of WASP-12b.
The groups of red points are the transit intervals, which are centered on the predicted transit
times and extend for 4 transit durations.}
\label{fig:wasp-12-time-series}
\end{figure*}

We counted the number of available data points during each interval.
If it was smaller than 75\% of the expected number
based on the transit duration and the nominal cadence
(2, 10, or 30~min), we omitted the
interval from further consideration. This removed transits
for which there were significant gaps in the time coverage
due to observing interruptions or other problems.
Figure \ref{fig:wasp-12-time-series} shows WASP-12 as an illustrative example.
The groups of red points are the transit intervals.

To remove photometric trends on timescales longer than
the transit duration, whether due to stellar variability or instrumental systematics, we fitted a polynomial to the data in each transit interval
that were obtained outside of the transit. The degree of the polynomial
was either 1, 2, or 3, with the decision based on minimizing the
Bayesian Information Criterion (BIC; \citealt{bic}),
$BIC=\chi^2 +k\log n$, where $k$ is the number of free parameters,
and $n$ is the number of data points.
In most cases, a linear model (degree 1) was selected.
We then ``rectified'' the data for each transit interval
by dividing the flux time series
by the polynomial function that had been fitted to the out-of-transit
data.

Next, we phase-folded the light curve, initially using the orbital period from TEPCat,
and subsequently using the updated orbital period from the first
iteration of our analysis.
We fitted a \cite{2002ApJ...580L.171M} model, for which
the free parameters were
the planet-to-star radius ratio $R_p/R_\star$,
the orbit-to-star radius ratio $a/R_\star$,
the impact parameter $b \equiv a \cos I/R_\star$ (where
$I$ is the orbital inclination),
the mid-transit time, and either one or two limb-darkening coefficients.
We used the BIC to decide whether to fit a linear or quadratic
limb-darkening profile; in almost all cases, a linear model
was chosen.
In a few cases with 30-minute sampling and an especially low SNR, we used a linear law with a fixed coefficient of 0.6.

For given values of $a/R_\star$ and $b$, the transit timescale
is proportional to the orbital period
(see, e.g., Equation~19 of \citealt{2010arXiv1001.2010W}). To establish this timescale,
we used the orbital period reported by TEPCat.
Although our aim was to
improve on the precision in the period, the uncertainty in the
previously reported period was already so small as to be irrelevant
for the determination of the basic transit parameters.
We sought the best-fitting model parameters by minimizing the usual $\chi^2$ statistic using the Nelder-Mead optimizer within the \textsc{lmfit} Python library. To minimize
the effects of occasional outliers,
we removed the data points lying more than 5-$\sigma$ away from
the best-fitting model value, where $\sigma$ is
the median absolute deviation (MAD) of the residuals.
We iterated this process until there were no more 5-$\sigma$ outliers.

\begin{figure*}

\begin{center}
\includegraphics[width=1\textwidth]{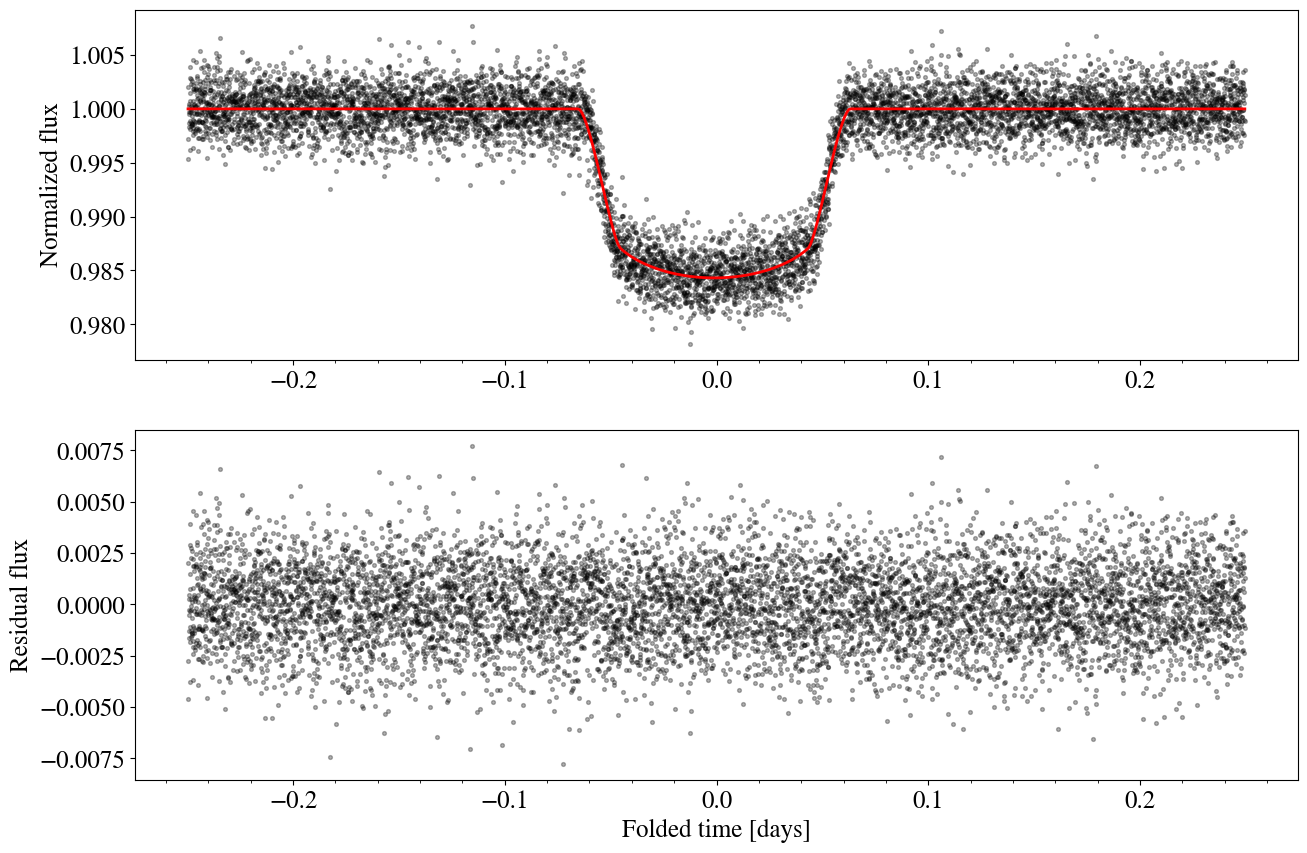}
\end{center}
\caption{Folded light curve of WASP-12b.}
\label{fig:folded-lc}
\end{figure*}

\begin{figure*}[hbt!]

\begin{center}
\includegraphics[trim=0cm 6.0cm 0cm 0cm, width=0.7\textwidth]{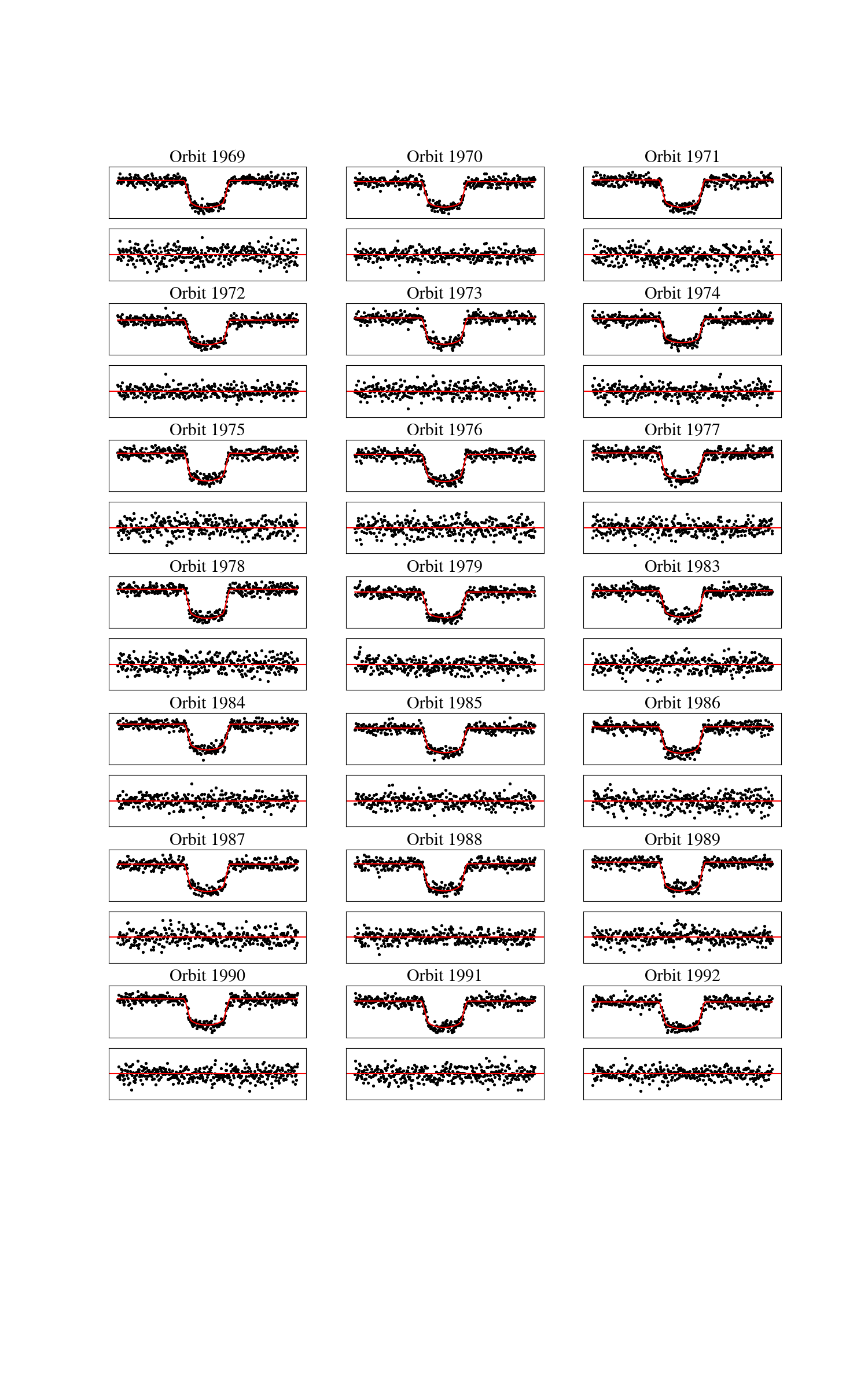}
\end{center}
\caption{Individual transits of WASP-12b. The red curves are the best-fitting models.}
\label{fig:individual-transits}
\end{figure*}

Having established good values for the parameters describing the shape
of the transit light curve, we proceeded to analyze each individual transit rather than the phase-folded light curve.
We held fixed the parameters $R_{p}/R_\star$, $a/R_\star$, $b$, and
the parameter or parameters of the limb-darkening law.
The only free parameters were the transit time and
the coefficients of the detrending polynomial. 
Figure~\ref{fig:folded-lc} illustrates
the phase-folded light curve of WASP-12 along with the best-fit transit model, and
Figure~\ref{fig:individual-transits} shows the individual transits.

To identify and reject transits for which the data quality was unusually
poor, we calculated the MAD of the photometric data for each transit
as well as the MAD of all the photometric data for multiple transits
that were observed in a given sector.
If a particular transit had MAD greater than 1.5 times the overall MAD, we omitted the transit
from further consideration. (This test was only performed when three or more
transits were observed per sector.)
If any transits were rejected in this step,
we reconstructed the phase-folded light curve,
rederived the geometric transit parameters and
refitted the individual transits.

We employed a Markov Chain Monte Carlo (MCMC) method to estimate the transit times and their uncertainties. The transition distribution was proportional to $\exp(-\chi^2/2)$ with
\begin{equation}
\chi^2 = \sum_{i=1}^{N} \left(
\frac{f_{{\rm obs},i} - f_{{\rm calc},i}}{\sigma_i}
\right)^2,
\end{equation}
where $f_{{\rm obs},i}$ is the observed flux at time $t_i$,
$f_{{\rm calc}, i}$ is the calculated flux at time $t_i$,
and $\sigma_i$ is the standard deviation of the out-of-transit data.

The 30-minute data were handled separately. When fitting models to the 30-minute
data, we computed the model with 2-minute sampling
and averaged the model into 30-minute bins before
comparing the model with the data.
We noticed that the 30-minute data produced by the QLP
tended to have higher levels of systematic effects, including
occasional bursts of apparent flux variability of unknown origin.
We visually identified about 10 cases in which these systematic effects
occurred during transits, and omitted these cases from our subsequent
analysis.  In a few other cases, the SNR of the TESS
transit detections proved to be too low for reliable transit
timing; these, too, were omitted from further consideration. In particular, we omitted K2-237 (Sector 12), MASCARA-4 (Sector 10, 11), KPS-1 (Sector 14), and WASP-8 (Sector 29) light curves from our analysis.

\section{Literature search} \label{sec:literature}

 A semi-automated procedure was used to search the extensive literature on transiting planets to find all the previously reported times for all the systems in our sample. First, we queried the NASA Astrophysics Data System database by Object
for each planet in our sample. We downloaded the {\tt\!.bib} file
associated with the search result, which is a text file that includes citation information for all
of the papers that refer to the object in question.
Some targets have alternative names; we tried using all
the names known to us.
We extracted the arXiv ID of each paper from the ``eprint'' entries in the {\tt\!.bib} file.

The next step was to download the {\tt\!.tex} file containing
the body of text for each paper.
We constructed the URL of the arXiv version of the paper
by concatenating
{\tt https://arxiv.org/e-print/} and the arXiv ID (1911.09131, for example). We used the \textsc{requests} Python library to download the {\tt\!.tex} files. Afterwards, we used the \textsc{tarfile} Python library to
expand the tar packages that arrived from arXiv.
Usually, in addition to the main {\tt\!.tex} file,
each package included supplementary materials
(such as tables of transit times).
There were only a few cases of papers for
which there was no arXiv version; these were handled separately by downloading
the published papers.

After the {\tt\!.tex} files had been downloaded, our code searched
the source files and noted all cases in which (i) the name of the
target appears at least once, and (ii) at least one number appears
that is greater than $2{,}000{,}000$ (allowing for the possible
presence of commas), which is plausibly a timestamp expressed
as a Julian Date.  For these cases, we used the arXiv Python
library to download the PDF versions of the manuscripts. In some papers,
transit times were reported as an abbreviated Julian Date, such as JD~$-$~2{,}400{,}000. Our automated procedure captured these cases as long
as the constant that was subtracted was mentioned somewhere in the paper and was greater than 2{,}000{,}000.
Any papers that did not report the offset might have been missed, although we expect such cases to be rare
based on our manual investigation
of cases of special interest and our experience with the literature.

At this stage, we needed to inspect the manuscripts to collect
all of the transit times pertaining to the object in question, and exert
judgment over whether the times had been measured well and reported
clearly.  We did not attempt to automate this step because of a host
of practical issues, among them the need to establish whether
the time system was HJD$_{\rm UTC}$, BJD$_{\rm TDB}$, or something else.
The basic rules we used at this stage were:
\begin{itemize}

\item When necessary,
we converted the reported times to the BJD$_{\rm TDB}$ system
using the time utilities code of \cite{Eastman_2010}.
Whenever the time system was reported as HJD without
further specification, we assumed the time system was HJD$_{\rm UTC}$
and converted the time accordingly. When the
label was BJD without further specification,
we assumed the time system was BJD$_{\rm TDB}$.

\item When the reported uncertainties were asymmetric, with
different upper and lower error bars,
for simplicity
we adopted the larger error bar and assumed the uncertainty
is symmetric.

\item In some cases, the only reported transit times 
were based on fitting data from multiple transits
and assuming a constant orbital period.
While not ideal for our purposes,
we did include these transit times
in our compilation and
identified them as such 
in Table~\ref{tbl:transit-times}.

\end{itemize}

For 57 out of the 382 planets in our sample,
no TESS data were available and there were no transit times in the literature
beyond the ephemeris from the paper reporting the discovery of the object.
Obviously, we were not able to improve on our knowledge
of these systems, but for completeness we still included
them in Table~\ref{tbl:transit-times}, reproducing the ephemerides provided by TEPCat.

We considered using additional transit times from the Exoplanet
Transit Database, which compiles data from both professional and amateur
astronomers, though we ultimately decided to use only the
transit times reported in the professional literature.

\begin{figure*}
\begin{center}
\includegraphics[width=1\textwidth]{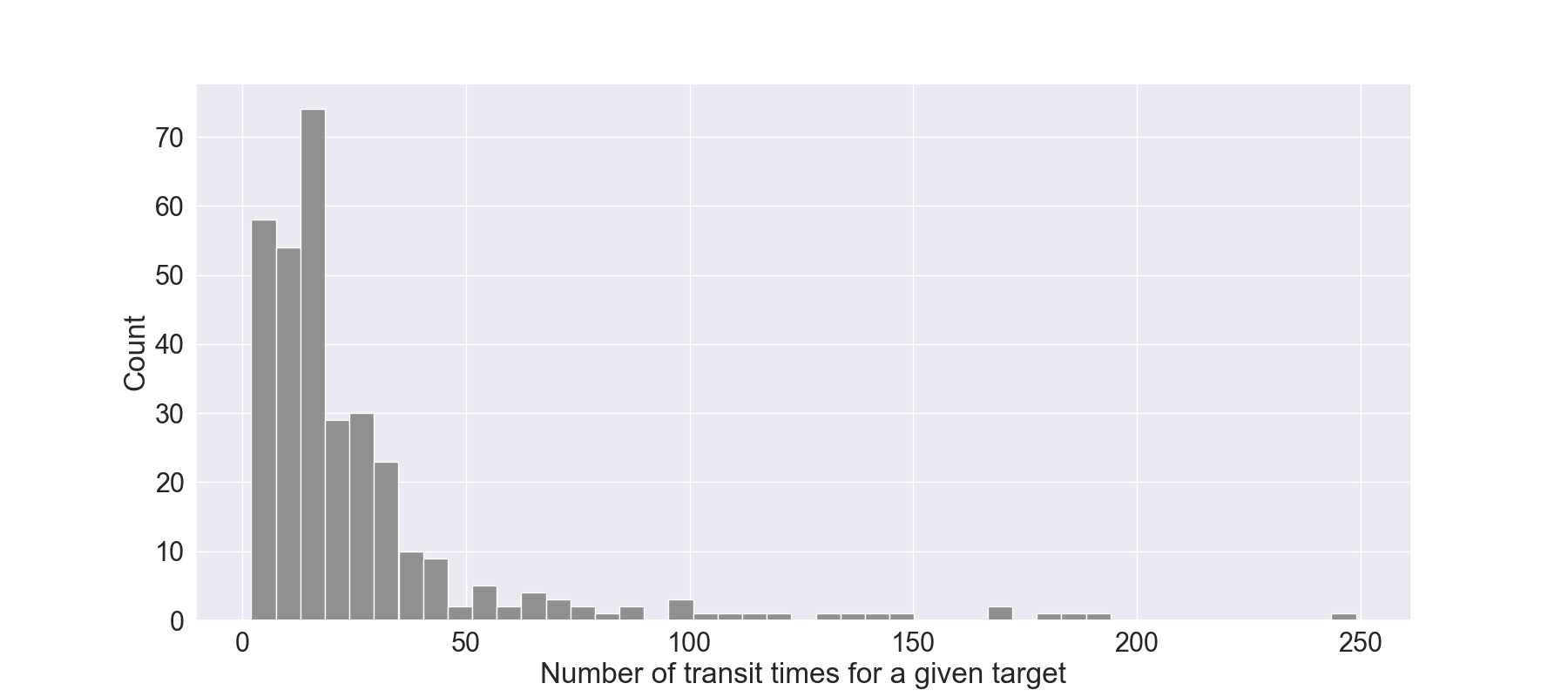}
\end{center}
\caption{Histogram of the number of transit times available for a given target after sigma-clipping has been applied when fitting a linear model to timing data.}
\label{fig:hist_num_points}
\end{figure*}

Table \ref{tbl:transit-times} provides 8{,}667 transit times for 382 planets, based on new TESS observations as well as
data from about 500 papers in the literature. For completeness, this table includes the cases for which only TESS data are available,
for which only earlier data are available, and for which both types of data are available.
Figure \ref{fig:hist_num_points} shows the distribution
of the number of transit times available for a given system. The median number
of transit times per planet is 16.

\section{Transit Timing Analysis} \label{sec:tt}

\subsection{TESS transit times}

First, we considered only the TESS transit times.
For each system with TESS data, we
used least-squares linear regression to fit a constant-period model to the TESS
transit times:
\begin{equation}
T_n = T_0 + nP,
\end{equation}
where $P$ is the period and $n$ is the number of transits that have occurred since a designated reference transit.
To reduce the covariance between the fitted values of $T_0$ and $P$,
we chose the reference transit to be the observed transit
closest to the median time of all the observed transits.

Assuming the model is correct and that the deviations between the data and the model
are independently and normally
distributed zero-mean random variables with 
variance $\sigma^2_i$, the $\chi^2$ statistic
should be distributed as
\begin{equation}
f(\chi^2) = \frac{(\chi^2)^{N_{\rm dof}/2-1}}{2^{N_{\rm dof}/2}\,\Gamma(N_{\rm dof}/2)} \exp\left(-\chi^2/2\right),
\end{equation}
where $N_{\rm dof}$
is the number of degrees of freedom
and $\Gamma(N_{\rm dof}/2)$ is the gamma function.
The $\chi^2$ distribution has a mean of $N_{\rm dof}$ and a 
variance of $2N_{\rm dof}$. 
Here, $N_{\rm dof}$ is the number of data points minus two.
Figure \ref{fig:chi-square-ndof} (left panel)
shows the minimum value of $\chi^2$ for each system
as a function of the number of degrees of freedom for that
system. 
In almost all cases, $\chi^2$ falls between the 2.5\% and 97.5 \% levels of the chi-square distribution.
We interpret this result as evidence that most systems in our sample do not exhibit short-term transit timing variations, and as a validation of our method for assigning uncertainties.

\subsection{All available transit times}

We also fitted the constant-period model to the entire collection
of transit times for each system, from both TESS and the literature.
We had two goals:
to identify systems showing evidence for period changes,
and to provide values and uncertainties for $T_0$ and $P$ that are
appropriate for predicting future transit times.

At first, we fitted the constant-period model taking all the data and
the reported uncertainties at face value, and recorded the minimum
value of $\chi^2$.
Any system for which the minimum value of $\chi^2$ was found to
exceed $N_{\rm dof} + 3\sqrt{2N_{\rm dof}}$
was flagged as a candidate for transit timing variations.
However, there were many cases in this category for which only one or
two data points deviated strongly from the best-fit model, at seemingly
random epochs.
Because the previously reported transit times are from heterogeneous sources
and are based on data sets with highly variable quality --- and likely
include some outright errors in the data analysis or in our
transcription --- we needed to make decisions about how to handle
large outliers in the timing data.

First, to avoid missing anything obviously interesting,
we visually inspected all of the timing
residuals regardless of $\chi^2$,
to look for any patterns in the residuals such as a gradual trend or
an oscillation, as opposed to isolated outliers.
To allow users of the database to perform similar visual
inspection, we have provided
online\footnote{\url{https://transit-timing.github.io}}
the entire library of figures showing
the light curves and the timing residuals for each system.
The cases that were flagged in this step
as potentially interesting are discussed below.

\begin{figure*}
\begin{center}
\includegraphics[width=1\textwidth]{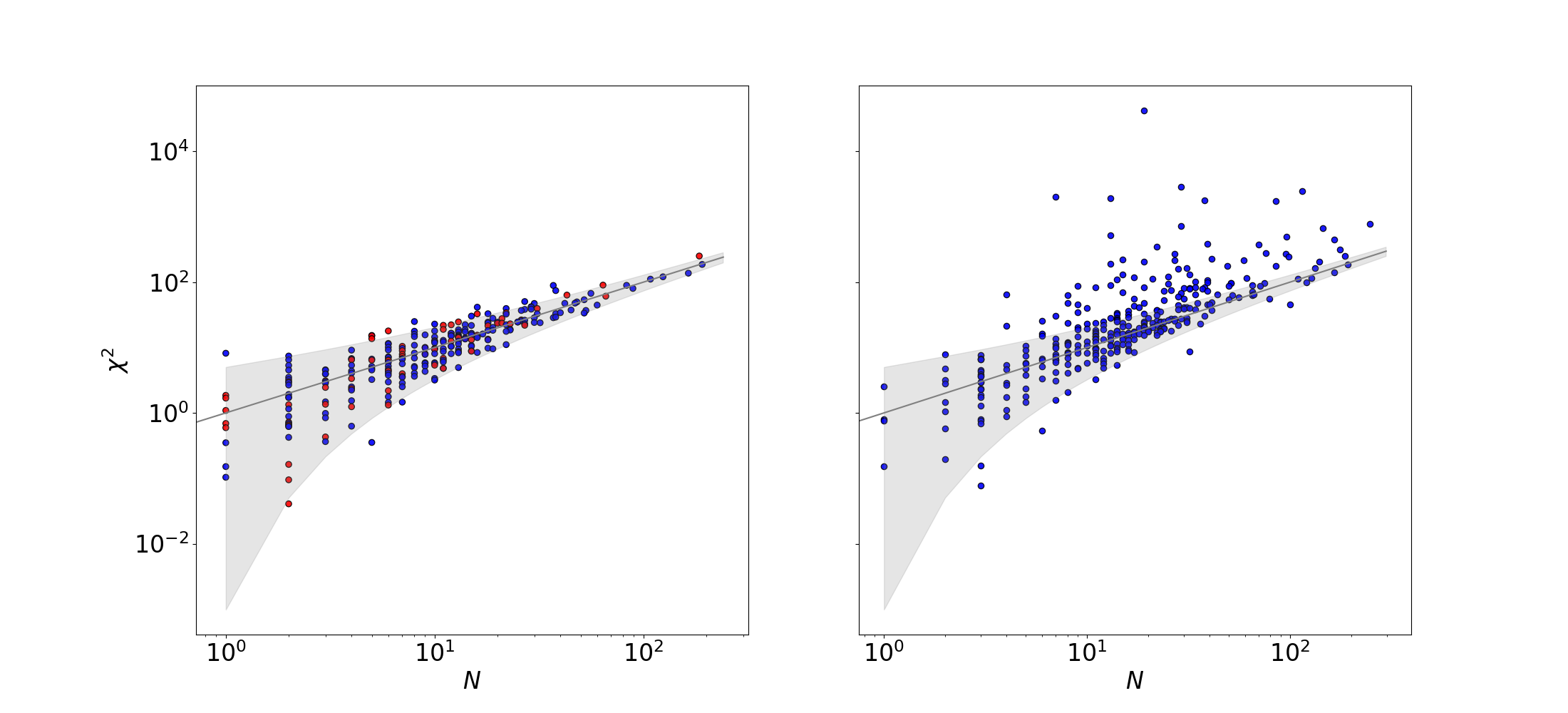}
\end{center}
\caption{Distribution of $\chi^2$ based on constant-period
models fitted to the transit times of all the planets in our sample.
The gray shaded area encompasses the region between the $2.5\%$ and $97.5\%$ levels
of the chi-square distribution.
(Left) Using TESS data only.
Red points are cases for which at least some of the
data had 30-minute sampling.
(Right) Using TESS data and previously reported data from the literature.}
\label{fig:chi-square-ndof}
\end{figure*}

Next, out of a general concern
about systematic errors in light-curve fitting
and the neglect of time-correlated noise,
we repeated the constant-period fit after
imposing a minimum uncertainty in each transit time of 0.0003 days (26 seconds),
and attaching zero weight to isolated $>$5-$\sigma$ outliers.
Figure \ref{fig:chi-square-ndof} (right panel) shows the
resulting distribution of
$\chi^2$ versus $N_{\rm dof}$.
Even after these adjustments, only about three-quarters of the systems
had a minimum $\chi^2$ between the 2.5\% and 97.5 \% levels
of the theoretical distribution, indicating that either there are genuine
transit timing variations or that the uncertainties in some of the data points have been underestimated.

For the purpose of assigning realistic uncertainties to future predicted
transit times, we re-fitted the data after adding another parameter $\sigma_0$
representing the level of either systematic timing errors or transit timing
variations.
This term was added in quadrature
to the reported uncertainty for each of the transit times whenever we had $\chi^2 > N_{\rm dof}$.
The value of $\sigma_0$ for each system was determined by the condition
$\chi^2 = N_{\rm dof}$.
For each target, we report
the value of $\chi^2$ obtained with $\sigma_0=0$,
as well as the value of $\sigma_0$ that
gives $\chi^2 = N_{\rm dof}$.
Table \ref{tbl:ephemerides} gives the results.

For the 57 targets for which no TESS data were available and multiple
independently-derived transit times have not been reported in the literature,
Table \ref{tbl:ephemerides} simply reproduces the ephemerides reported in TEPCat, for the convenience of having the
information for many systems in one place. 

For three targets -- WASP-161\,b, XO-6\,b, and HAT-P-19\,b -- there were some major outliers in the timing diagrams based on transit times reported in the literature that caused us to suspect problems with the data. After analyzing TESS data of XO-6\,b, \cite{2020AJ....160..249R} have also arrived at a similar conclusion that there might be unknown timing errors in its literature data. Table \ref{tbl:transit-times} includes all of the available transit times, including these outliers. However, Table  \ref{tbl:ephemerides} reports ephemerides for WASP-161\,b and XO-6\,b derived from TESS data only. In the case of HAT-P-19\,b, we excluded a single outlier ($T_0 = 2458699.53657 \pm 0.00043$ BJD$_{\rm TDB}$  reported in \citet{2020MNRAS.496.4174B}) when deriving the ephemeris.  

\begin{deluxetable*}{cccccccccc}
\tablecaption{Ephemerides.}
\tablewidth{0pt}
\tablehead{
\colhead{System} & \colhead{$N_{\rm data}$} &  \colhead{$\chi^2$} &
 \colhead{ $\sigma_0$} & \colhead{$P$ (days)} & \colhead{Uncertainty (days)} & \colhead{$T_0$}
  & \colhead{Uncertainty (days)}   & \colhead{ Time system }   & \colhead{Reference}
 }
\startdata
CoRoT-01 & 67 & 63.83 & 0 &
1.508968772 & $8.3 \times 10^{-8}$ &
2456268.99119 & 0.00011 & BJD$_{\rm TDB}$ & This work 
\enddata
\tablecomments{$N_{\rm data}$ is the number of transit times that
were fitted, $\chi^2$ is the sum of squared residuals when using
the timing uncertainties reported in the literature, $\sigma_0$
is the error term that must be added in quadrature to the
timing uncertainties of each of the transit times to achieve $\chi^2=N_{\rm dof}$,
and $T_0$ and $P$ are the timing parameters and uncertainties
obtained by fitting the data after adding $\sigma_0$ in quadrature
to the timing uncertainties of each of the transit times.
Only a portion of this table is shown here to demonstrate its form and content.
A machine-readable version of the full table can be accessed from the electronic version of this work.}
\label{tbl:ephemerides}
\end{deluxetable*}

\section{Cases of special interest}\label{sec:special}

\subsection{Features in light curves}

\begin{figure*}
\begin{center}
\includegraphics[width=1\textwidth]{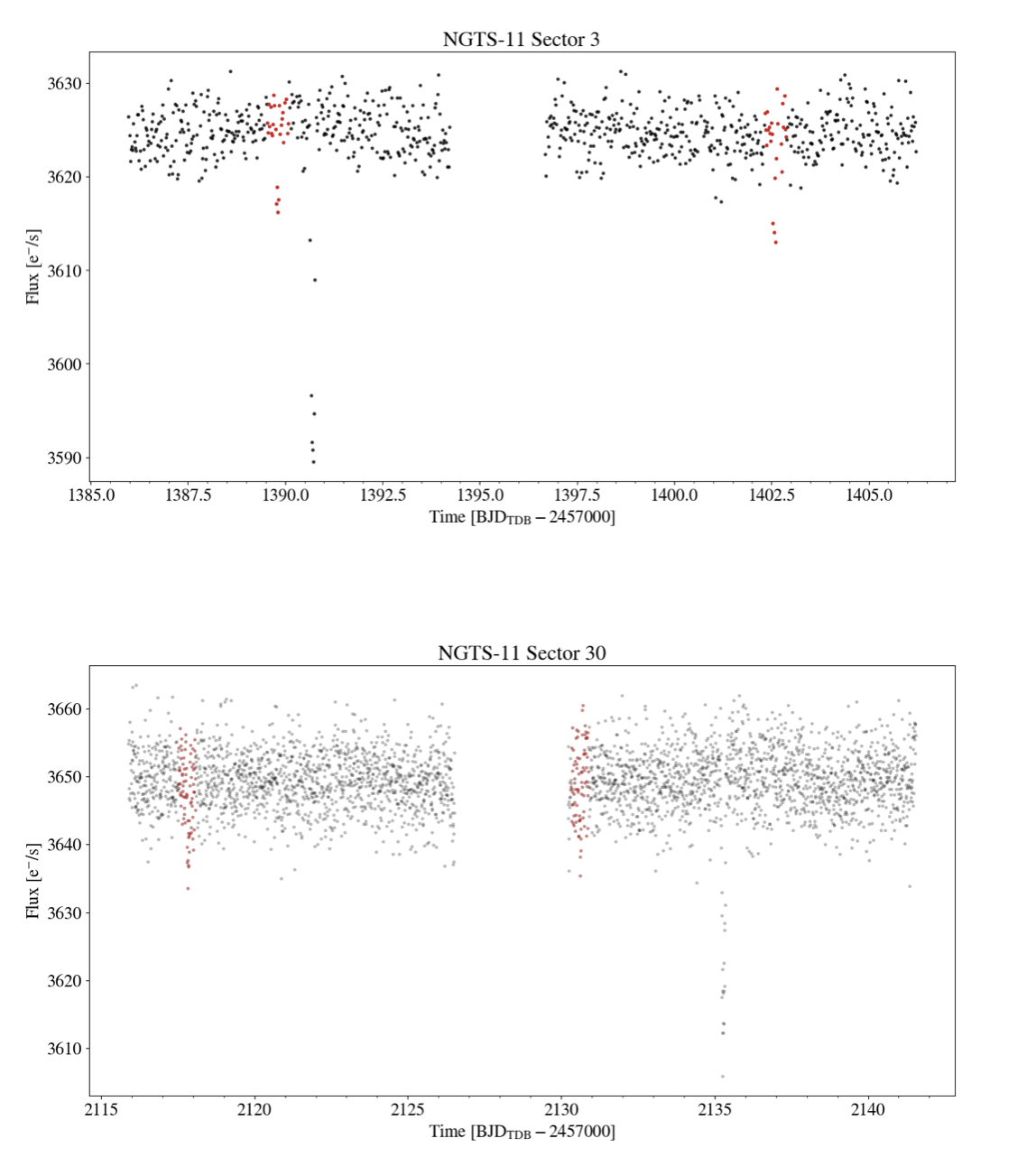}
\end{center}
\caption{NGTS-11 TESS light curves. The transits of NGTS-11\,c are shown in red.}
\label{fig:ngts-11-lc}
\end{figure*}

\begin{figure*}
\begin{center}
\includegraphics[width=0.8\textwidth]{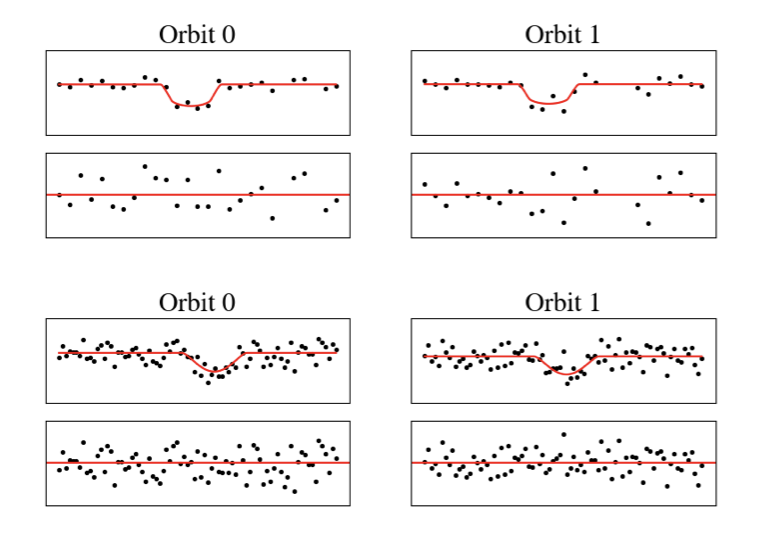}
\end{center}
\caption{NGTS-11\,c individual transits (upper panel: TESS sector 3; lower panel: TESS sector 30).}
\label{fig:ngts-11-individual-transits}
\end{figure*}

\begin{figure*}
\begin{center}
\includegraphics[width=0.92\textwidth]{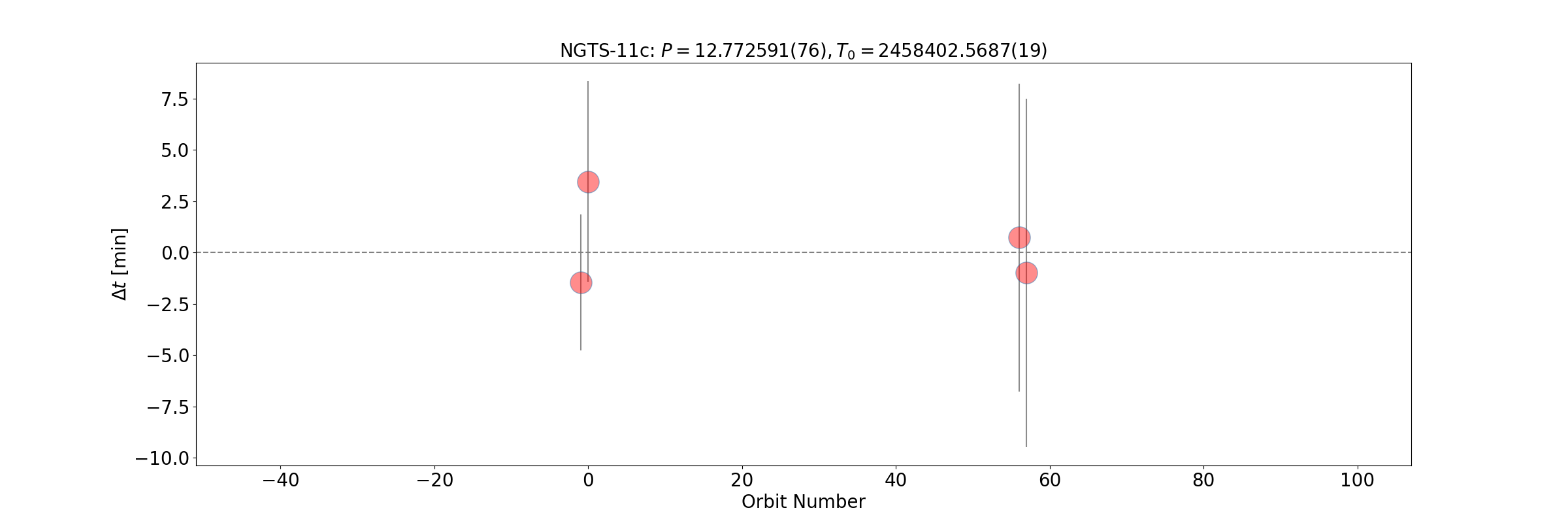}
\end{center}
\caption{Timing residuals of NGTS-11\,c. The points are based on \textit{TESS} data.}
\label{fig:ngts-11c-timing-residuals}
\end{figure*}

Although our goal was transit timing, we also looked out
for any unexpected features in the TESS light curves.
For the following systems, we noticed additional fading
events besides those due to the known transiting planets:
\begin{itemize}

\item {\bf NGTS-11} shows four dips, all with an amplitude of about 0.3\%,
at TESS-JDs 1389.8, 1402.57, 2117.83, and 2130.6 (sectors 3 and 30).\footnote{The TESS-JD is
defined as BJD minus 2{,}457{,}000.} See Figures~\ref{fig:ngts-11-lc}, ~\ref{fig:ngts-11-individual-transits}, and ~\ref{fig:ngts-11c-timing-residuals}.
These characteristics are compatible with an orbital period of 12.77 days
and a planet radius of 4.7\,$R_\oplus$, assuming the stellar
radius is 0.832\,$R_\odot$ \citep{Gill+2020}.
The previously reported planet in this system
has a period of 35.5 days and a radius of 9.2\,$R_\oplus$.
After noticing these extra dips, we also saw that this object appears
in the latest
list of TESS Objects of Interest with the identifier TOI~1847.02.

\item {\bf HATS-25} shows a dip near TESS-JD 2315 (sector 37).
We checked the TESS images for this system.
The extra dip is also seen
in a light curve based on
seemingly blank pixels in the image. We hypothesize
that the extra dip is due to a problem
with the automated subtraction of a model for the background
light in the image.

\item {\bf WASP-77} shows a closely spaced pair of dips
between TESS-JD 2151.0 and 2151.5 (sector 37). Inspection
of the images also showed that these dips
appear in light curves constructed from pixels
far from the star. Thus, the dips almost certainly
do not belong to WASP-77.

\item The light curve of {\bf WASP-83}
shows a brightening event at TESS-JD 1590.575.
A similar brightening event can be seen in the light curve
of every pixel in rows 1368 and 1369. The time of the event
becomes progressively later with column number. Clearly, these
events are unrelated to WASP-83. Perhaps
they are the effect of a moving object (such as an asteroid)
that passed through the field or interfered with background
subtraction.

\end{itemize}

For {\bf KELT-9} and {\bf MASCARA-4},
the transit signals do not 
have the usual mirror symmetry around the time of minimum light.
This was already known from prior data \citep{Ahlers+2020a, Ahlers+2020b}.
The reason for the asymmetry is the combination of two
properties:
(i) the stars are rapidly rotating, leading to the latitudinal
variation in emergent intensity known as gravity darkening,
and (ii) the planets' orbits are misaligned with respect
to the stellar equator.

The TESS data for {\bf HAT-P-11} show occasional anomalies during
transits characteristic of spot crossings, when the planet
moves in front of a relatively dark patch on the star's
photosphere. Such anomalies were seen in profusion
in the NASA Kepler data for this object by \cite{SanchisOjedaWinn2011} and
others.

The TESS light curves for the following systems showed rapid
oscillations that are characteristic of stellar
pulsations: {\bf HAT-P-32}, {\bf HAT-P-49}, {\bf KELT-8},
{\bf WASP-33}, {\bf WASP-178},
and possibly {\bf KELT-24} and {\bf WASP-7}.
Because we did not
attempt to model or filter out these oscillations,
the transit times for these systems are subject
to unusually large systematic uncertainties and the
reported ephemerides should be used with caution.
For {\bf WASP-178}, in particular, the
oscillations are due to blending
of the star with the nearby
pulsating star ASASSN-V J150908.07-424254.1.
We did not include TESS transit times for this target in our work.

\subsection{Long-term period changes}

To look for evidence of long-term period changes in each system,
we fitted a model in which the period is changing at a steady rate:
\begin{equation}
T_n = T_0 + nP + \frac{n^2}{2P}\frac{dP}{dt}.\\
\end{equation}
and flagged any cases for which $dP/dt$ was found to
be at least 3-$\sigma$ away from zero.
These cases are discussed in this section, along with a few
other notable systems.

\begin{figure*}
 \begin{center}
\includegraphics[width=1\textwidth]{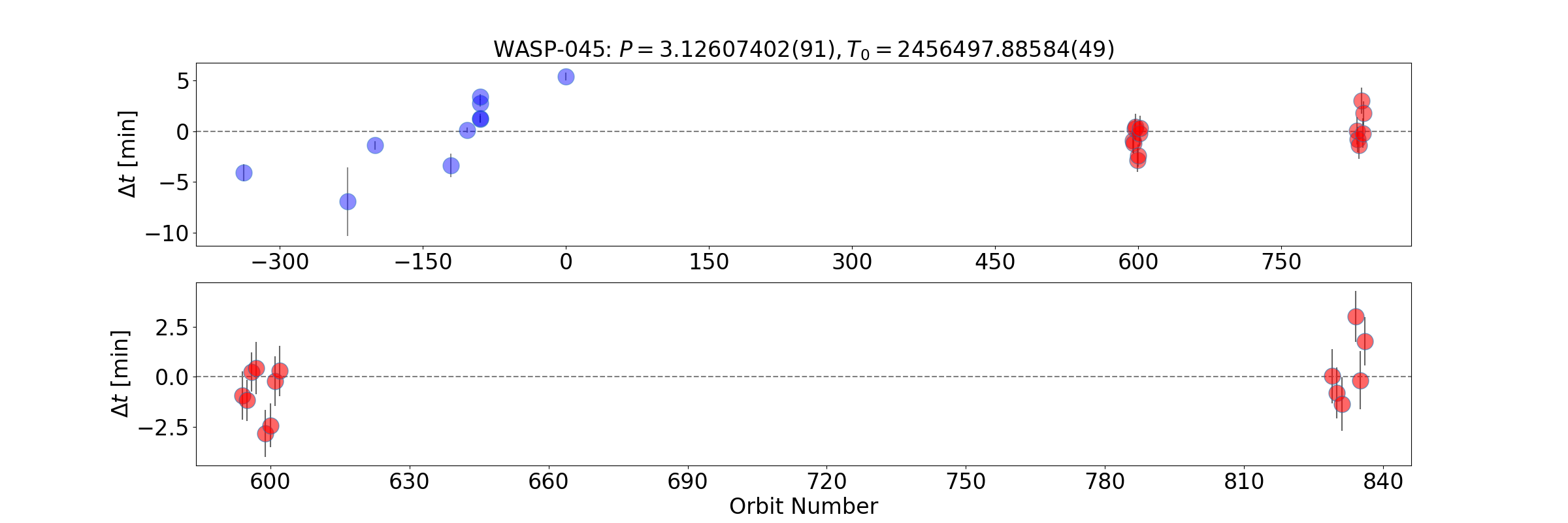}
\end{center}
\caption{Top panel: timing residuals of WASP-45\,b. Blue points are based on previously reported
transit times drawn from the literature. Red points are based on TESS data. Bottom panel: close-up view of the \textit{TESS} data points.}
\label{fig:wasp-045_o_c}
\end{figure*}

\begin{figure*}
\begin{center}
\includegraphics[width=1\textwidth]{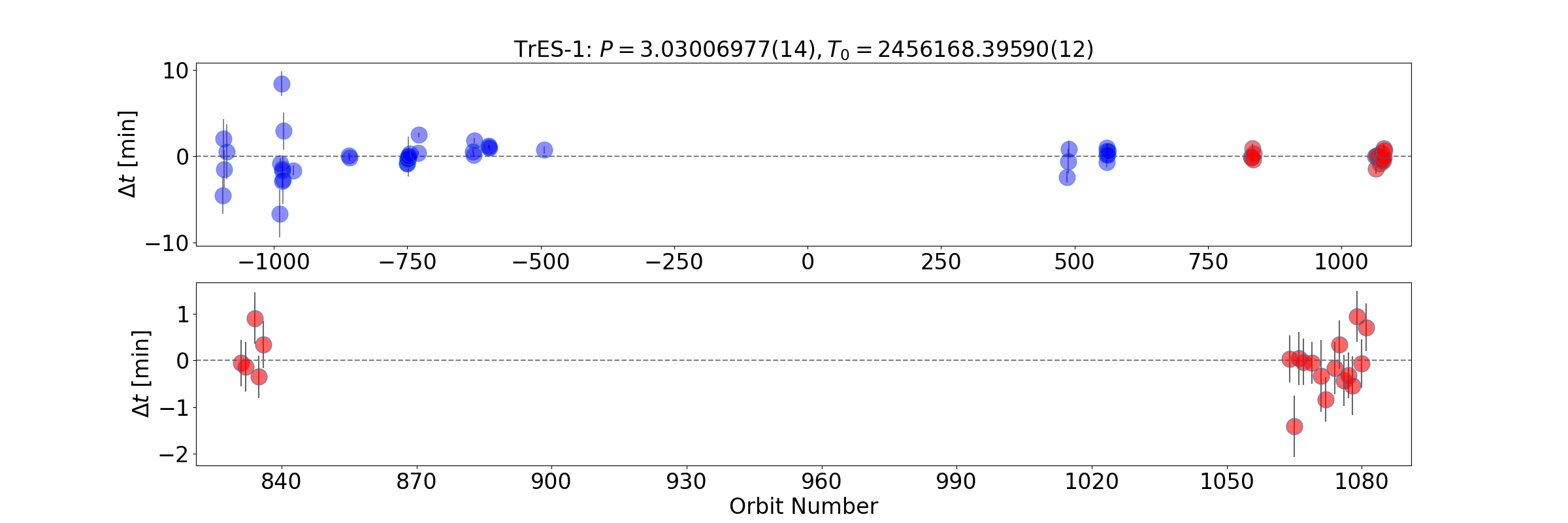}
\end{center}
\caption{Top panel: timing residuals of TrES-1\,b. Blue points are based on previously reported
transit times drawn from the literature. Red points are based on TESS data. Bottom panel: close-up view of the \textit{TESS}  data points.}
\label{fig:tres-1-0_c}
\end{figure*}

\begin{figure*}
\begin{center}
\includegraphics[width=1\textwidth]{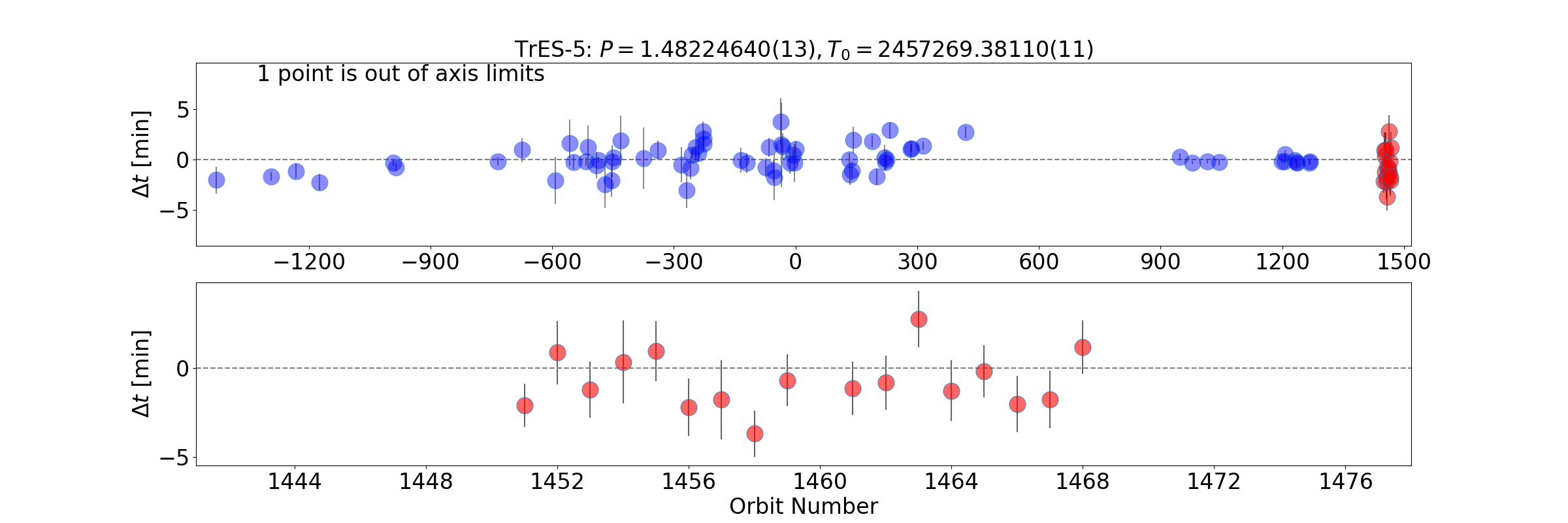}
\end{center}
\caption{Top panel: timing residuals of TrES-5\,b. Blue points are based on previously reported
transit times drawn from the literature. Red points are based on TESS data. Bottom panel: close-up view of the \textit{TESS}  data points.}
\label{fig:tres-5-0_c}
\end{figure*}

\begin{figure*}
\begin{center}
\includegraphics[width=1\textwidth]{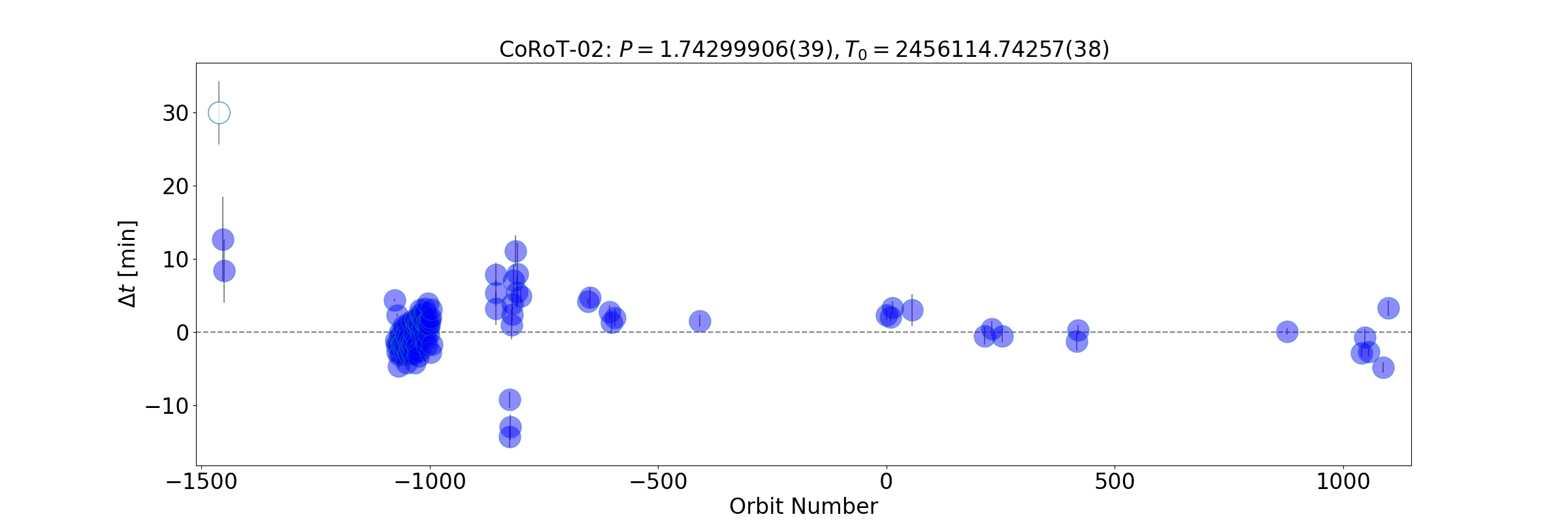}
\end{center}
\caption{Timing residuals of CoRoT-2\,b. Blue points are based on previously reported
transit times drawn from the literature. The white point is a $5-\sigma$ outlier that was not included when performing the fits. }
\label{fig:corot-02}
\end{figure*}

\begin{figure*}
\begin{center}
\includegraphics[width=1\textwidth]{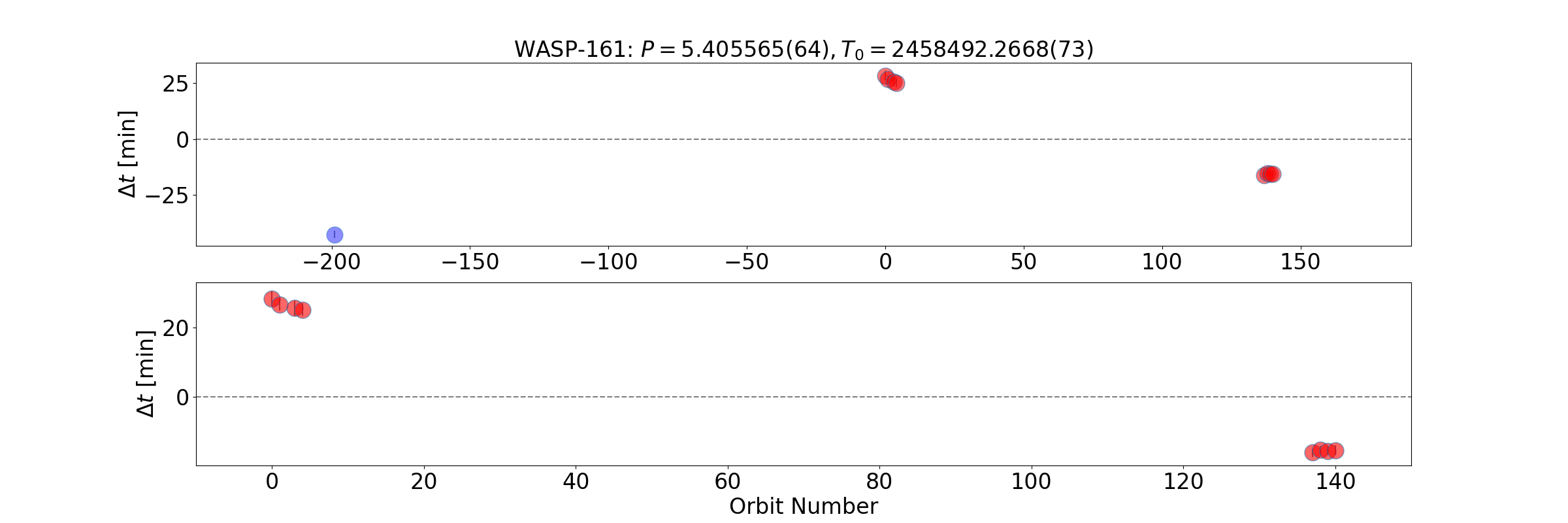}
\end{center}
\caption{Top panel: timing residuals of WASP-161\,b. The blue point is based on a previously reported
transit time drawn from the literature. Red points are based on TESS data. Bottom panel: close-up view of the \textit{TESS}  data points.}
\label{fig:wasp-161-o_c}
\end{figure*}

\begin{figure*}
\begin{center}
\includegraphics[width=1\textwidth]{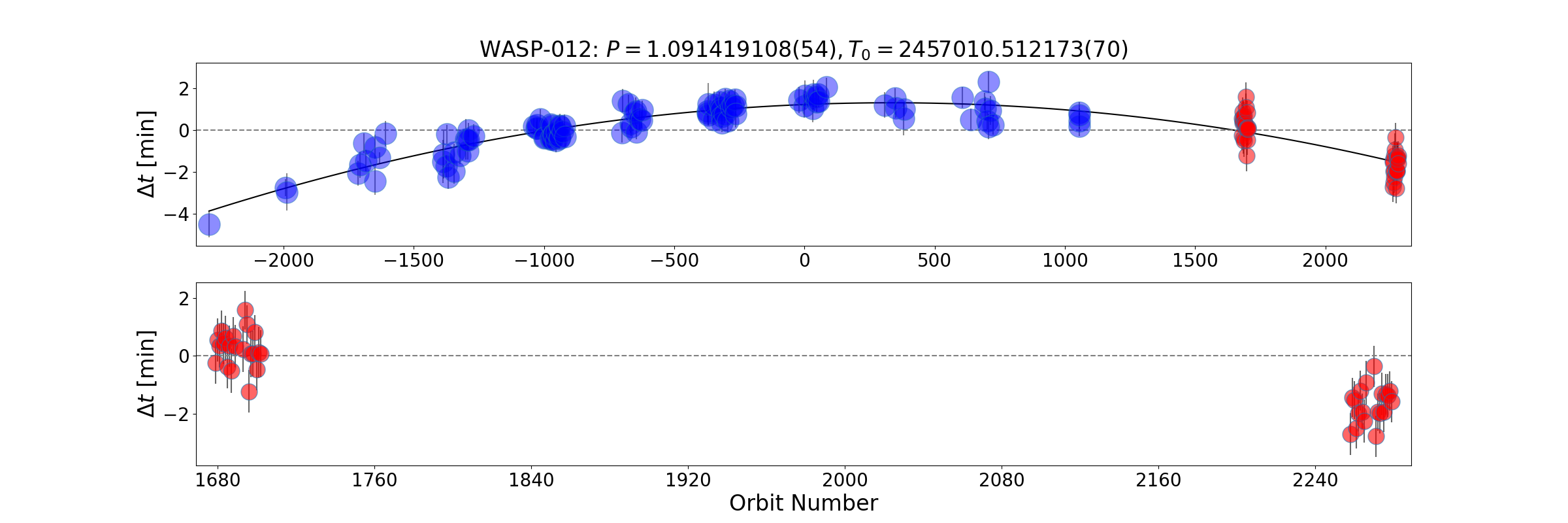}
\end{center}
\caption{Top panel: timing residuals of WASP-12\,b. Blue points are based on previously reported
transit times drawn from the literature. Red points are based on TESS data. The black curve shows the residuals of the best-fitting model in which the period changes uniformly with time. Bottom panel: close-up view of the \textit{TESS} data points.}
\label{fig:wasp-12-o_c}
\end{figure*}

\begin{figure*}
\begin{center}
\includegraphics[width=1\textwidth]{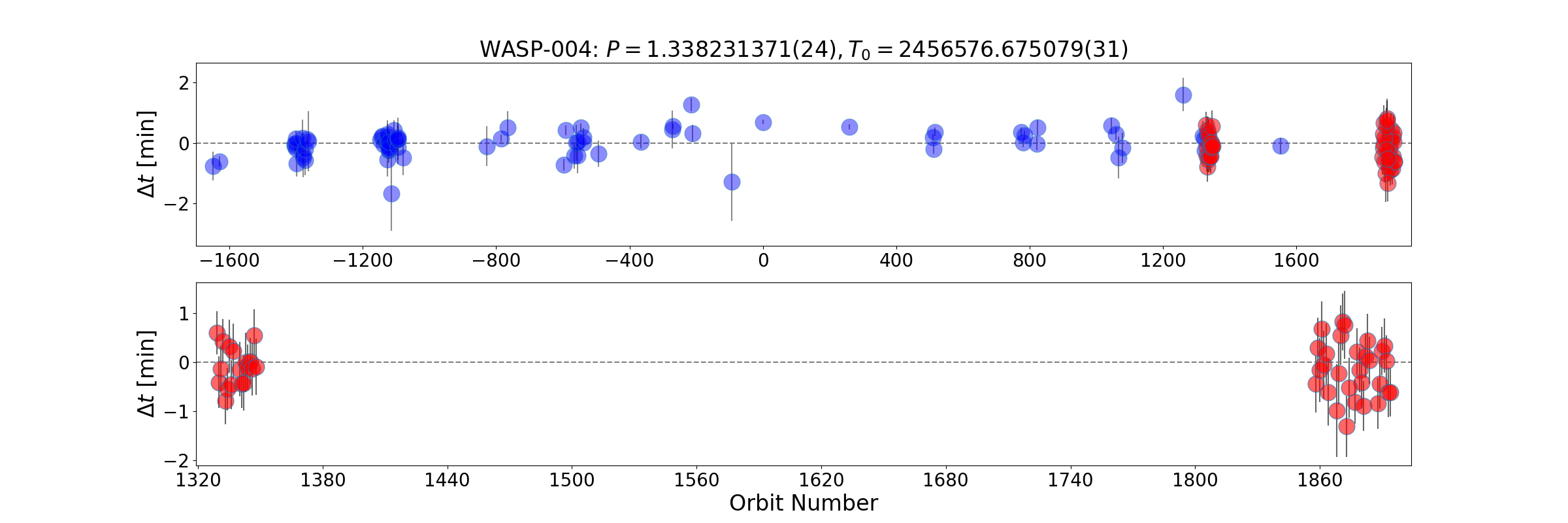}
\end{center}
\caption{Top panel: timing residuals of WASP-4\,b. Blue points are based on previously reported
transit times drawn from the literature. Red points are based on TESS data. Bottom panel: close-up view of the \textit{TESS} data points.}
\label{fig:wasp-4-o_c}
\end{figure*}

\begin{figure*}
\begin{center}
\includegraphics[width=1\textwidth]{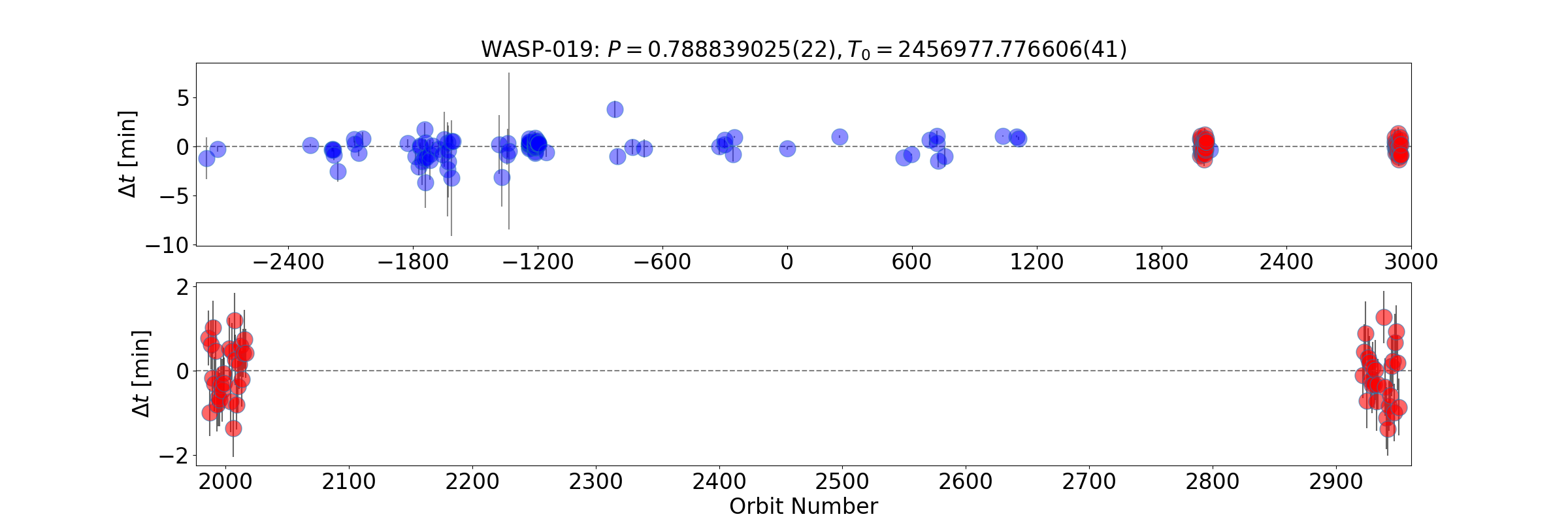}
\end{center}
\caption{Top panel: timing residuals of WASP-19\,b. Blue points are based on previously reported
transit times drawn from the literature. Red points are based on TESS data. Bottom panel: close-up view of the \textit{TESS} data points.}
\label{fig:wasp-019-o_c}
\end{figure*}

\begin{itemize}

\item {\bf WASP-12\,b} is a $1.5\,M_{Jup}$ planet in a 1.1-day orbit around a star that appears to be a late-F main-sequence star, but could also be a subgiant (\citealt{2009ApJ...693.1920H,2017ApJ...849L..11W}).
The transit period
is slowly decreasing with time \citep{Maciejewski+2016, 2017AJ....154....4P}.
The interval between occultations (secondary eclipses) is also decreasing
with time \citep{Yee_2019}. Thus, the 
orbit appears to be shrinking, possibly due to tidal orbital decay.
TESS transit timing for this system was performed by \cite{Turner+2021AJ},
who found $dP/dt = -32.53\pm 1.62$~ms/yr. We found 
$dP/dt = -30.27\pm 1.11$~ms/yr (see Figure~\ref{fig:wasp-12-o_c}).

\item \textbf{WASP-4\,b} is a $1.2\,M_{Jup}$ planet on a 1.3-day orbit around
a G7V star \citep{2008ApJ...675L.113W}. The transit period was found to be decreasing in an earlier
analysis of TESS data \citep{Bouma_2019}.
Most or all of the observed change in the transit
period is due to the long-term acceleration of the star toward
the Sun, probably due to the gravitational pull of a wide-orbiting
companion \citep{Bouma+2020}. We found $dP/dt = -5.81\pm 1.58$~ms/yr
(see Figure~\ref{fig:wasp-4-o_c}) which is compatible with the result reported by \cite{Bouma_2019}
and is based on an additional sector of TESS data.

\item \textbf{WASP-45\,b} is a $1.0\,M_{Jup}$ planet in a 3.1-day orbit around a K2V star \citep{Anderson_2012}.
The best-fit model gives $dP/dt = -262.57 \pm 28.35$ ms/yr (see Figure \ref{fig:wasp-045_o_c}).
However, the pattern of timing residuals is not obviously quadratic,
and the early times show scatter far in excess of the formal uncertainties.
By themselves, the TESS data show no evidence for a period change. We consider
this a mediocre candidate for a period change; more data are needed.

\item \textbf{CoRoT-2\,b} is a $3.3\,M_{Jup}$ planet in a 1.7-day orbit around a G7V star (\citealt{2008A&A...482L..21A,2008A&A...482L..25B}). The star is active and rapidly rotating, with spots covering $\sim$10\% of its visible surface \citep{Guillot_2011}. 
TESS data are not yet available, but our compilation of previous measurements
gave $dP/dt = -103.76 \pm 6.33$ ms/yr (see Figure~\ref{fig:corot-02}).
This, too, does not yet seem like a compelling candidate for a period change, because of the large scatter
around the quadratic trend and because the starspots can cause errors in the
determination of transit times.

\item \textbf{TrES-1\,b} is a $0.76\,M_{Jup}$ planet in a 3.0-day orbit around a K0V star \citep{Alonso_2004}.
 
Including the TESS data, we found $dP/dt = -18.36 \pm 3.73$ ms/yr. Visually, the timing residuals do not show any particular pattern (see Figure~\ref{fig:tres-1-0_c}),
and as usual, the data drawn from the literature include sporadic outliers, so it is difficult
to take the 3-$\sigma$ detection too seriously. Still, this system might be worth further attention.

\item \textbf{TrES-5\,b} is a $1.8\,M_{Jup}$ planet in a 1.5-day orbit \citep{tres5}.
Including the TESS data, we found $dP/dt = -17.47 \pm 3.79$ ms/yr. The timing residuals are shown in Figure~\ref{fig:tres-5-0_c}. A recent study by \cite{maciejewski2021revisiting} concluded that the orbital period of the planet could be varying on a long timescale. The authors found that the most likely explanation of the observations is the line-of-sight acceleration of the system's center-of-mass due to the orbital motion induced by a massive, wide-orbiting companion. This system is worth watching further.

\item \textbf{WASP-161\,b} is a $2.5\,M_{Jup}$ planet in a 5.4-day orbit around an F6V star \citep{Barkaoui_2019}.
The only previously reported timing data was from
\cite{Barkaoui_2019}, who computed an ephemeris but did not provide individual
transit times. The reference epoch of their ephemeris was
$2457416.5289 \pm 0.0011$~HJD, which we converted to BJD$_{\rm TDB}$ before
including it in our timing study.
However, using the TESS data alone, we would have predicted the transit referenced by
\cite{Barkaoui_2019} to have occurred more than an hour earlier (see Figure \ref{fig:wasp-161-o_c}).
The best-fit model has $dP/dt = -15.5 \pm 0.35$~{\it seconds} per year (not ms/yr).
Because the timing residuals have a single outlier, an error in determining or reporting the transit
time seems at least as plausible as a genuine timing variation of such a large amplitude.

\item \textbf{WASP-99\,b} is a $2.8\,M_{Jup}$ planet in a 5.8-day orbit around an F8V star \citep{Hellier_2014}. 
The best-fit model has $dP/dt = -305.85\pm 104.98$ ms/yr, but this
is another case in which the literature provide only a single transit time, based on a fit
to data from multiple transits and the assumption of a constant period.
This makes any conclusions vulnerable to any problems with that single datum.

\item \textbf{WASP-19\,b} is a $1.1\,M_{\rm Jup}$ planet in a 0.78-day orbit around a G8V star \citep{2010ApJ...708..224H}.
It is predicted to be one of the most favorable targets in the search for evidence of tidal orbital decay. 
The timing residuals are shown in Figure~\ref{fig:wasp-019-o_c}.  
The best-fit quadratic model gives
$dP/dt = -3.54 \pm 1.18$~ms/yr, which just barely makes our 3-$\sigma$ criterion. Ordinarily, we would
not call attention to such a weak detection.
However, given that it is based on voluminous and generally high-quality data,
and given the prior expectation that the orbit should be decaying especially rapidly,
this is most certainly a system worth watching carefully. We hope that TESS keeps revisiting
this system every few years.

\item \textbf{XO-3\,b} is a 13\,$M_{\rm Jup}$ planet on a 3.2-day orbit around an F5 star,
with an eccentric and misaligned orbit \citep{JohnsKrull+2008, Hebrard+2010}.
The period derivative for this system was found to be
$dP/dt = -182.08\pm 12.96$~ms/yr.   
However, the earliest timing residuals show a lot of scatter,
characteristic of ground-based light curves with small telescopes (see Figure~\ref{fig:xo-3-o_c}).
Nevertheless, this system is notable because the combination of the TESS data
and two transit times measured precisely with data from the {\it Spitzer Space Telescope} \citep{10.1088/0004-637x/794/2/134}
appear to be incompatible with a constant period. Additional visits to this system by
TESS should clarify the situation.

\end{itemize}

\subsection{Short-term transit timing variations}

There are only two convincing cases of short-term transit timing
variations in our sample. 
The first case is \textbf{Kepler-448}, an F star with a transiting 1.2\,$R_{\rm Jup}$ planet on an 18-day orbit,
for which transit timing variations revealed an outer giant planet
on an eccentric orbit \citep{10.3847/1538-3881/aa7aeb}. TESS has only observed a single transit,
so the new data do not add much information.
The second case is \textbf{WASP-148}, a G dwarf with two known planets,
one of which is a transiting 0.29\,$M_{\rm Jup}$ planet
on an 8.8-day orbit, and the other of which is a non-transiting
planet with a minimum mass of 0.4\,$M_{\rm Jup}$ and a period
of 34.5 days. The transit timing variations of the inner
planet, including the TESS data, have
been analyzed by \cite{Maciejewski2020ANM}.

\begin{figure*}
\begin{center}
\includegraphics[width=1\textwidth]{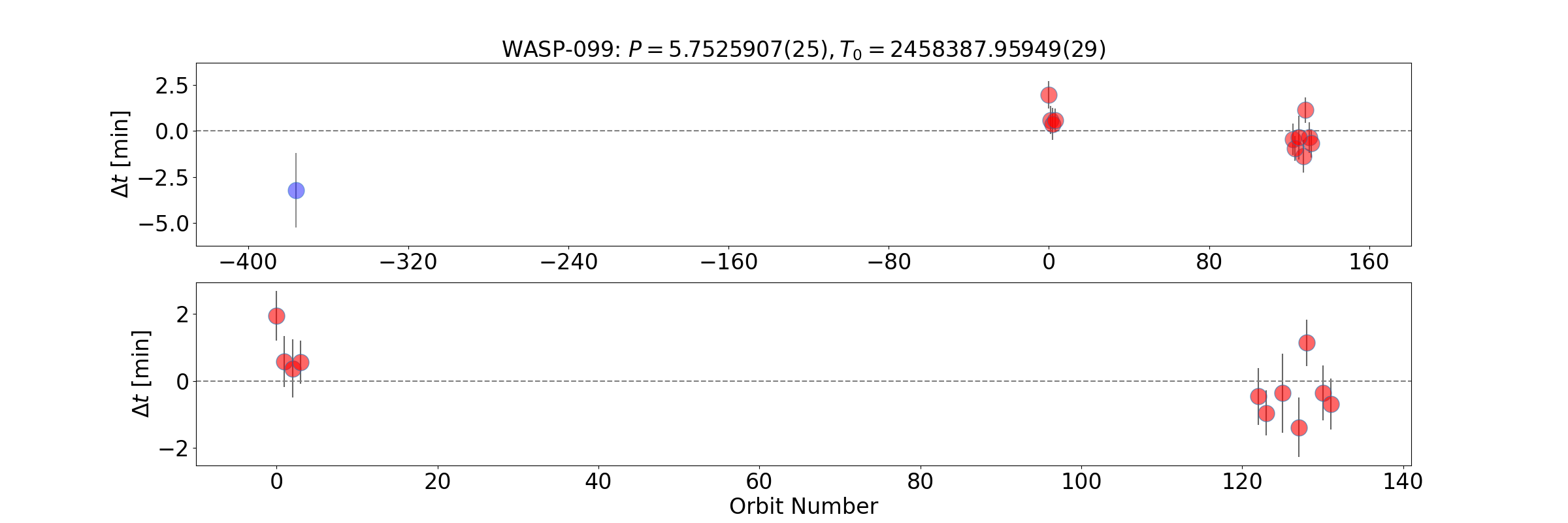}
\end{center}
\caption{Top panel: timing residuals of WASP-99\,b. The blue point is based on a previously reported
transit time drawn from the literature. Red points are based on TESS data. Bottom panel: close-up view of the \textit{TESS} data points.}
\label{fig:wasp-099-o_c}
\end{figure*}

\begin{figure*}
\begin{center}
\includegraphics[width=1\textwidth]{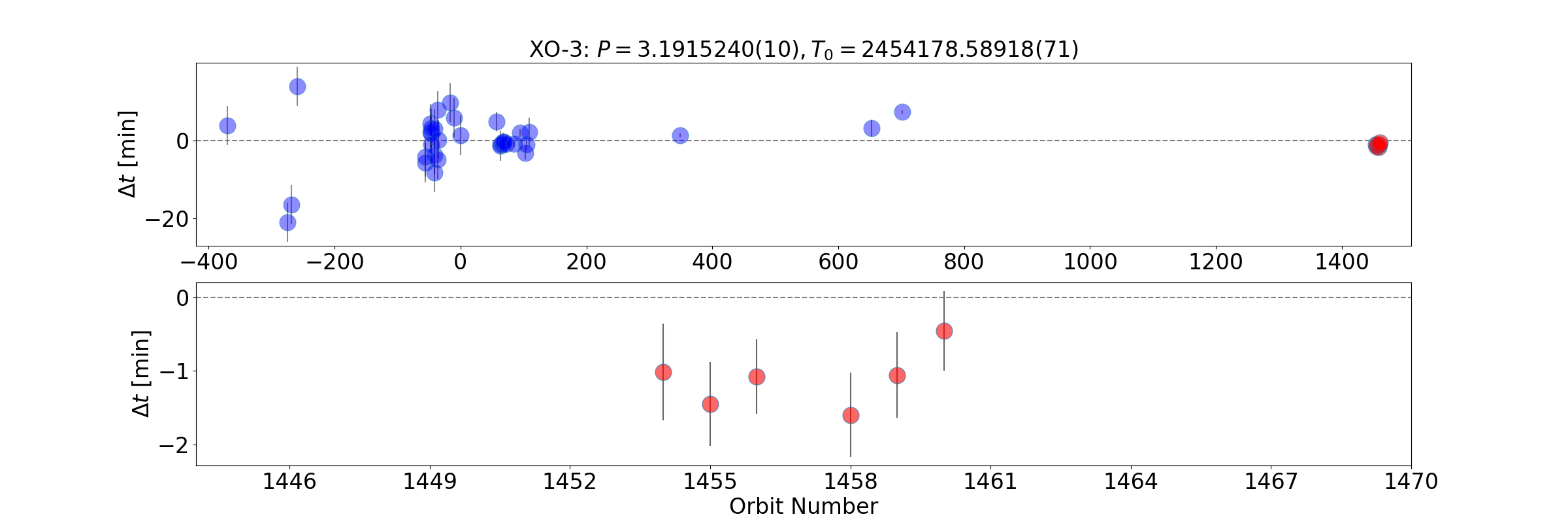}
\end{center}
\caption{Top panel: timing residuals of XO-3\,b. Blue points are based on previously reported
transit times drawn from the literature. Red points are based on TESS data. Bottom panel: close-up view of the \textit{TESS} data points.}
\label{fig:xo-3-o_c}
\end{figure*}

\section{Summary and Discussion} \label{sec:summary}

\begin{figure*}
\begin{center}
\includegraphics[width=1\textwidth]{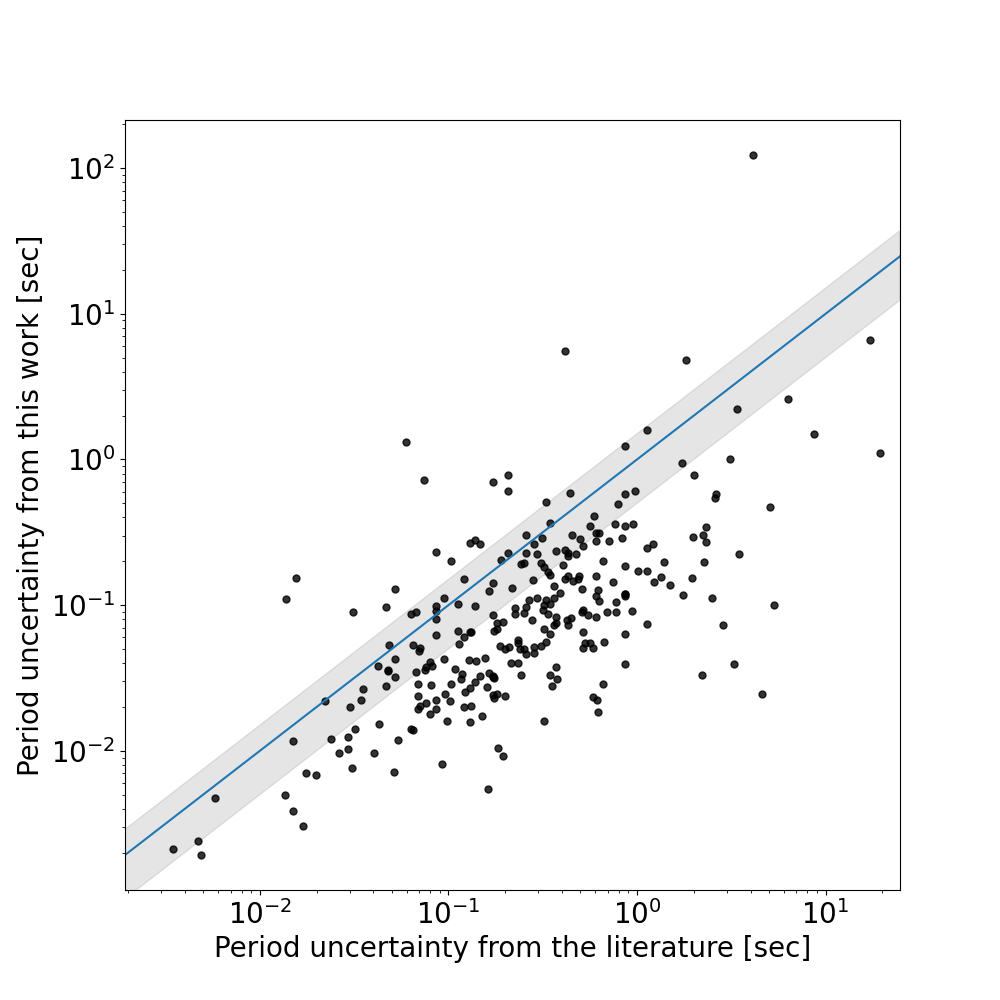}
\end{center}
\caption{Orbital period error based on fitting literature data only \textit{versus} orbital period error derived in our work. Grey area encapsulates the region between 0.5 times the identity line and 1.5 times the identity line. The identity line is shown in blue.}
\label{fig:period_uncs}
\end{figure*}

By analyzing TESS data and compiling data from the literature,
we improved the precision with which the orbital period is known
for 246 planets (Figure \ref{fig:period_uncs}).  
We hope that this database of transit times will be useful to anyone interested in observing these systems in the future,
as well as anyone
searching for long-term period changes,
or constraining population-level properties from
the incidence (or lack) of transit timing variations.
Nevertheless, any automated procedure applied to hundreds of systems is bound to fumble in some cases. Users of the database are
encouraged to inspect the light curves and
timing residual diagrams provided
online (\url{https://transit-timing.github.io}) before trusting
the ephemerides to plan observations with
expensive facilities such as the {\it James Webb Space Telescope.}

To facilitate future studies of this nature,
we encourage astronomers to report transit times
in the BJD$_{\rm TDB}$ time system, and to report
measurements of {\it individual} transit times
in addition to the ephemerides derived from fitting all
the transit times.

After this work was nearly complete, we became aware of two other
groups, \cite{Kokori+2021} and \cite{Shan+2021}, who have also
been performing transit timing {\it en masse}.   Since these works
are not yet published, we refrained from a detailed comparison
between our numerical results and theirs, but we can make some general
comparisons.

\cite{Kokori+2021} reported ephemerides for 180 planets
based on data from the literature as well as new
photometric data collected by the ExoClock network, which is a community of amateur astronomers. This team did not utilize TESS data in their study.

\cite{Shan+2021} identified 31 hot Jupiters for which an ephemeris
taken from the literature failed to predict the TESS transit times
to within the statistical uncertainty.
This is different from our approach, which was based on analyzing
all available transit times whenever possible,
rather than relying on a previously
reported ephemeris.
Their list includes WASP-161 and WASP-99, which we discussed
in Section~\ref{sec:special}. We did not
consider these to be
compelling cases for period changes
because of the paucity of data points.
Apart from those two cases, we did not flag any of the
same systems as candidates for period changes.

\section{Code}
The code is available on github: \url{https://github.com/transit-timing/tt}. 
The code is archived  in Zenodo: \cite{https://doi.org/10.5281/zenodo.5904270}.

\section{Aacknowledgements}
{We are grateful to Filip Walter for compiling and providing the Exoplanet
Transit Database in a convenient format, although we ultimately did not include
those data in our analyses. We thank Sarah Millholland, Vadim Krushinsky, and
Luke Bouma for helpful comments on the draft of this paper. We would also like
to thank Betsy Pu for her help with the development of our project's website.
This work was supported by the Heising-Simons Foundation and the NASA TESS
mission. This work relied on the NASA Astrophysics Data System and the arXiv.
We also acknowledge the use of the Extrasolar Planets Encyclopedia at
exoplanet.eu \citep{2011A+A...532A..79S}; the NASA Exoplanet Archive, which is
operated by the California Institute of Technology, under contract with the
National Aeronautics and Space Administration under the Exoplanet Exploration
Program \citep{https://doi.org/10.26133/nea12}. The authors are pleased to
acknowledge that the work reported in this paper was substantially performed
using the Princeton Research Computing resources at Princeton University which
is a consortium of groups led by the Princeton Institute for Computational
Science and Engineering and Office of Information Technology's Research
Computing.
}

\appendix

\begin{deluxetable*}{ccccccc}
\tablecaption{Transit times.}
\tablehead{
\colhead{System} & \colhead{Orbit number} &  \colhead{Transit time} &
 \colhead{ Uncertainty (days)}  & \colhead{Time System}
  & \colhead{\#}   & \colhead{ Reference }}
\startdata
CoRoT-01 & -1412 & 2454138.32761 &	0.00047 & BJD &	1 & 2009A\&A...506..369B \\
\enddata
\tablecomments{\# column indicates if the transit time is based on an individual transit event or a fit of multiple transits simultaneously assuming constant period. 
Only a portion of this table is shown here to demonstrate its form and content. A machine-readable version of the full table can be accessed from the electronic version of this work.}
\label{tbl:transit-times}
\end{deluxetable*}

Table \ref{tbl:transit-times} provides a database of transit times for each of the 382 planets included in this work. The table includes the cases for which only TESS data are available,
for which only earlier data are available, and for which both types of data are available.
 The references in Table 4 are:
\citet{2016ApJ...819...27F}; 
\citet{2020AJ....159..109M}; 
\citet{2010A&A...519A..98B}; 
\citet{2007A&A...476L..13B}; 
\citet{2015PASP..127..143R}; 
\citet{2018MNRAS.474.2334D}; 
\citet{2017AJ....153...97O}; 
\citet{2010ApJ...719.1796B}; 
\citet{2006A&A...459..249G}; 
\citet{2009ApJ...691.1145S}; 
\citet{2019A&A...628A...9C}; 
\citet{2016PASP..128b4402S}; 
\citet{2014A&A...563A..41M}; 
\citet{2008A&A...488..763H}; 
\citet{2011ApJ...733..127S}; 
\citet{2019A&A...622A..81M}; 
\citet{2020AJ....160..233A}; 
\citet{2018A&A...610A..63D}; 
\citet{10.3847/1538-3881/aa7aeb}; 
\citet{2015AJ....150...85H}; 
\citet{2017AJ....154..122C}; 
\citet{2020AJ....159..204W}; 
\citet{2016A&A...585A.126W}; 
\citet{2018A&A...620A.142A}; 
\citet{2018ChA&A..42..101S}; 
\citet{2020AJ....159..150P}; 
\citet{2010MNRAS.408.1680S}; 
\citet{2017AJ....153...78C}; 
\citet{2013A&A...551A..80T}; 
\citet{2015A&A...577A.109M}; 
\citet{2018AJ....155...55B}; 
\citet{2011ApJ...735...24J}; 
\citet{2016AJ....152..127P}; 
\citet{2021AJ....162...18S}; 
\citet{2007ApJ...664.1185H}; 
\citet{2020MNRAS.496.4174B}; 
\citet{2011ApJ...742..116B}; 
\citet{2015ApJ...811...55M}; 
\citet{2014AJ....147...92W}; 
\citet{2020A&A...639A.130O}; 
\citet{2013MNRAS.436.2974G}; 
\citet{2019ApJS..240...13L}; 
\citet{2020AJ....160..111R}; 
\citet{2011A&A...527A...8S}; 
\citet{2011ApJ...742...59H}; 
\citet{2017MNRAS.469.1622M}; 
\citet{2018A&A...613A..41M}; 
\citet{2013AJ....145...68J}; 
\citet{2008ApJ...675L.113W}; 
\citet{2017MNRAS.471..650M}; 
\citet{2012A&A...544A..72L}; 
\citet{2012A&A...543L...5G}; 
\citet{2014A&A...568A.127M}; 
\citet{2013AJ....146..147M}; 
\citet{2011AJ....142...84F}; 
\citet{2010MNRAS.408.1494C}; 
\citet{2010AJ....139...53R}; 
\citet{2018PASP..130g4401B}; 
\citet{2011ApJ...741..102A}; 
\citet{2010ApJ...710.1724B}; 
\citet{2014A&A...563A.143D}; 
\citet{2009ApJ...700.1078G}; 
\citet{2020MNRAS.499.3139B}; 
\citet{2018AJ....155..112B}; 
\citet{2010ApJ...725.2017K}; 
\citet{2012ApJ...745...80Q}; 
\citet{2016arXiv160804225F}; 
\citet{2018AJ....155...83Z}; 
\citet{2010PASJ...62L..61N}; 
\citet{2011PASJ...63..301L}; 
\citet{2019EPSC...13..595E}; 
\citet{2014MNRAS.440.1982H}; 
\citet{2012A&A...547A..61S}; 
\citet{2018MNRAS.480..291S}; 
\citet{2009A&A...502..395W}; 
\citet{2020A&A...636A..98C}; 
\citet{10.1086/512721}; 
\citet{2016AcA....66...55M}; 
\citet{2018MNRAS.481.4261S}; 
\citet{2013A&A...549A..30N}; 
\citet{2006ApJ...652.1715H}; 
\citet{2006ApJ...651L..61O}; 
\citet{2013MNRAS.428..678T}; 
\citet{2020AcA....70..203M}; 
\citet{2011AJ....141..179C}; 
\citet{2015MNRAS.447..711S}; 
\citet{2009AJ....137.4834W}; 
\citet{2020AAS...23533707Z}; 
\citet{2011ApJ...728..138H}; 
\citet{2020AJ....160..109S}; 
\citet{2015A&A...581L...6D}; 
\citet{2021MNRAS.500.5420C}; 
\citet{2021AJ....161..108S}; 
\citet{2013A&A...552A.120S}; 
\citet{2010ApJ...720..337S}; 
\citet{2016A&A...591A..55M}; 
\citet{2011MNRAS.410.1631E}; 
\citet{2007ApJ...667L.195M}; 
\citet{2012A&A...545A..93S}; 
\citet{2020MNRAS.491.2760G}; 
\citet{2012AJ....143...81S}; 
\citet{2008PASJ...60L...1N}; 
\citet{2007A&A...465.1069P}; 
\citet{2017PASP..129j5001K}; 
\citet{2020ApJ...888L..15M}; 
\citet{2012AJ....144...19B}; 
\citet{2021MNRAS.501.5393L}; 
\citet{2010ApJ...709..159A}; 
\citet{2016AJ....151...89B}; 
\citet{2017AJ....154....4P}; 
\citet{2011AJ....142..115D}; 
\citet{10.1088/0004-637x/794/2/134}; 
\citet{2009PASP..121.1104J}; 
\citet{2017ApJ...836L..24W}; 
\citet{2019MNRAS.482..301L}; 
\citet{2018AJ....156..181W}; 
\citet{2013ApJ...772L...2D}; 
\citet{2017MNRAS.465..843M}; 
\citet{2014A&A...568A..81L}; 
\citet{2011PASP..123..555A}; 
\citet{2012NewA...17..438D}; 
\citet{2008MNRAS.387L...4A}; 
\citet{2018MNRAS.473.5126P}; 
\citet{2010ApJ...710...97C}; 
\citet{2011A&A...526A.130S}; 
\citet{2013ApJ...774...95D}; 
\citet{2016AJ....151...17J}; 
\citet{2015AJ....149..149B}; 
\citet{2020A&A...635A..60D}; 
\citet{2020AJ....159..267B}; 
\citet{2010MNRAS.407..507C}; 
\citet{2006ApJ...648.1228M}; 
\citet{2017AJ....153..263M}; 
\citet{10.3847/0004-637x/823/2/122}; 
\citet{2011ApJ...726...94C}; 
\citet{2017AJ....154...95H}; 
\citet{2012MNRAS.427.2757M}; 
\citet{2007PASJ...59..763N}; 
\citet{2018MNRAS.475.4467B}; 
\citet{2013A&A...549A.134H}; 
\citet{2013A&A...551A..73F}; 
\citet{2013ApJ...766...95L}; 
\citet{2019MNRAS.489.2478N}; 
\citet{2008ApJ...680.1450P}; 
\citet{2015A&A...583A.138M}; 
\citet{2012MNRAS.419.1248B}; 
\citet{10.1093/mnras/stv197}; 
\citet{2010IBVS.5919....1S}; 
\citet{2013ApJ...778..184J}; 
\citet{2010ApJ...715..458T}; 
\citet{2016MNRAS.459.4281K}; 
\citet{2007MNRAS.376.1296S}; 
\citet{2011MNRAS.416.2108A}; 
\citet{2009ApJ...696.1950B};  
\citet{2018PASP..130f4401W}; 
\citet{2009A&A...501..785G}; 
\citet{2014AJ....147...39C}; 
\citet{2016AJ....152..161D}; 
\citet{2013MNRAS.428.2645M}; 
\citet{2018AJ....155..100J}; 
\citet{2013A&A...553A..26N}; 
\citet{2018arXiv181209264A}; 
\citet{2020A&A...643A..45S}; 
\citet{2014A&A...570A..64S}; 
\citet{2009A&A...508.1011R}; 
\citet{2012MNRAS.426.1338S}; 
\citet{2020AJ....160..222J}; 
\citet{2017MNRAS.471.2743T}; 
\citet{2016PASP..128b4401S}; 
\citet{2021A&A...656A..88M}; 
\citet{2012MNRAS.422.1988A}; 
\citet{2017AJ....154...49W}; 
\citet{10.1051/0004-6361/201424744}; 
\citet{2016MNRAS.457.4205S}; 
\citet{2011AJ....142...95K}; 
\citet{2010MNRAS.401.2665P}; 
\citet{2014ApJ...785..148R}; 
\citet{2013RMxAA..49...71R}; 
\citet{10.1017/S1743921308026823}; 
\citet{2015MNRAS.451.4139R}; 
\citet{2018AJ....156..283E}; 
\citet{2011ApJ...726...52H}; 
\citet{2018AJ....156..124B}; 
\citet{2016arXiv160604556B}; 
\citet{2020AJ....160...47M}; 
\citet{2010ApJ...724..866K}; 
\citet{2012ApJ...748...22H}; 
\citet{2019MNRAS.486.2290O}; 
\citet{2009ApJ...703..752L}; 
\citet{2013A&A...558A..55M}; 
\citet{2019AJ....157..141T}; 
\citet{2018MNRAS.474..876K}; 
\citet{2012AJ....143...95L}; 
\citet{2012MNRAS.426..739H}; 
\citet{2017ApJ...848....9S}; 
\citet{2008ApJ...683.1076W}; 
\citet{2018A&A...612A..57T}; 
\citet{2012ApJ...750...84B}; 
\citet{2017A&A...606A..18L}; 
\citet{2019AJ....157...55H}; 
\citet{2009A&A...502..391S}; 
\citet{2017AJ....153..200A}; 
\citet{2014MNRAS.441..304S}; 
\citet{2013MNRAS.434.1300S}; 
\citet{2017A&A...599A...3L}; 
\citet{2012ApJ...746..111T}; 
\citet{2014AJ....148...29J}; 
\citet{2018AJ....156..216S}; 
\citet{2021ApJS..255...15W}; 
\citet{10.3847/1538-3881/ab8815}; 
\citet{2015MNRAS.450.3101B}; 
\citet{2015AJ....150...18H}; 
\citet{2010A&A...523A..84S}; 
\citet{2017PASP..129f4401R}; 
\citet{2011A&A...533A..88G}; 
\citet{2011ApJ...738...50A}; 
\citet{2018arXiv180907709A}; 
\citet{2011ApJ...730L..31H}; 
\citet{2014A&A...562A.126M}; 
\citet{2017Sci...356..628W}; 
\citet{2020ApJ...898L..11G}; 
\citet{2009MNRAS.396.1023S}; 
\citet{2012MNRAS.420.2580S}; 
\citet{2010ApJ...715..421T}; 
\citet{2011MNRAS.416.2593B}; 
\citet{2017AJ....154..194L}; 
\citet{2012ApJ...749..134H}; 
\citet{2015PASP..127..851J}; 
\citet{2015ApJ...811..122W}; 
\citet{2011ApJ...726....3N}; 
\citet{2020A&A...642A..54C}; 
\citet{2020ApJ...893L..29B}; 
\citet{2019AJ....157...31Z}; 
\citet{2009A&A...496..259G}; 
\citet{10.3847/1538-3881/aa9e4d}; 
\citet{2019MNRAS.483.1970B}; 
\citet{2017A&A...604A.110A}; 
\citet{2014AJ....147..161S}; 
\citet{2017AIPC.1815h0021P}; 
\citet{2015MNRAS.447..463N}; 
\citet{2014ApJ...790..108F}; 
\citet{2008ApJ...681..636I}; 
\citet{2017A&A...602A.107B}; 
\citet{2008A&A...492..603G}; 
\citet{2009A&A...506..359G}; 
\citet{2016ApJ...827...19F}; 
\citet{2020MNRAS.493..126M}; 
\citet{2013MNRAS.428.3680G}; 
\citet{2019MNRAS.483..824R}; 
\citet{2015A&A...580A..63M}; 
\citet{2015MNRAS.450.1760T}; 
\citet{2004ApJ...609L..37K}; 
\citet{10.1038/nature08856}; 
\citet{2015AJ....150..197H}; 
\citet{2013AJ....146..113B}; 
\citet{2018MNRAS.477L..21M}; 
\citet{2015MNRAS.446.1389P}; 
\citet{2020PASP..132a4401M}; 
\citet{2012A&A...539A..97S}; 
\citet{2007ApJ...657.1098W}; 
\citet{2015ApJ...814...66K}; 
\citet{2011AJ....141....8S}; 
\citet{2008A&A...485..871G}; 
\citet{2011A&A...534A..16A}; 
\citet{2016AJ....152..136Z}; 
\citet{2014AJ....147..144Z}; 
\citet{2005ApJ...626..523C}; 
\citet{2014MNRAS.437.1511B}; 
\citet{2015AJ....150..168H}; 
\citet{2011MNRAS.414.3023S}; 
\citet{2022MNRAS.509.5102S}; 
\citet{2012PASJ...64...97S}; 
\citet{2009ApJ...690L..89H}; 
\citet{2009A&A...506..369B}; 
\citet{2019AIPC.2178c0019B}; 
\citet{2014AJ....147..128H}; 
\citet{2020AJ....159..173H}; 
\citet{2016MNRAS.463.3276H}; 
\citet{2007AJ....134.1707W}; 
\citet{2009ApJ...704.1107L}; 
\citet{2015MNRAS.446.2428S}; 
\citet{2014MNRAS.444..776S}; 
\citet{2019AJ....157...43B}; 
\citet{2017AJ....154..237V}; 
\citet{2012AJ....144..139H}; 
\citet{2008ApJ...686.1331B}; 
\citet{2017AJ....153..178S}; 
\citet{2018IBVS.6243....1M}; 
\citet{2010A&A...517L...1Q}; 
\citet{2010ApJ...723L..60H}; 
\citet{2012A&A...539A.159N}; 
\citet{2011ApJ...734..109B}; 
\citet{2013MNRAS.434...46H}; 
\citet{2017MNRAS.472.3871T}; 
\citet{2011ApJ...733..116B}; 
\citet{2013A&A...552A...2L}; 
\citet{2014arXiv1410.3449A}; 
\citet{2019A&A...622A.172M}; 
\citet{2018AJ....155...52A}; 
\citet{2013ApJ...764....8K}; 
\citet{2020MNRAS.497.5182A}; 
\citet{2016arXiv160700322B}; 
\citet{2009ApJ...690.1393S}; 
\citet{2012ApJ...747...82C}; 
\citet{2017MNRAS.467.1714T}; 
\citet{2013MNRAS.432..693M}; 
\citet{2014arXiv1412.7761B}; 
\citet{2013A&A...552A..82G}; 
\citet{2016MNRAS.458.4025D}; 
\citet{2019AJ....158...63E}; 
\citet{2009ApJ...707..446B}; 
\citet{2016MNRAS.456..990C}; 
\citet{2011A&A...531A..60A}; 
\citet{10.3847/1538-3881/ab9f2d}; 
\citet{2019AJ....157..224A}; 
\citet{2016PASP..128f4401T}; 
\citet{2007AJ....133.1828W}; 
\citet{10.3847/1538-3881/aa8bb9}; 
\citet{2015A&A...577A..54C}; 
\citet{2017AJ....153...94C}; 
\citet{2020arXiv201107169S}; 
\citet{2009AJ....137.3826W}; 
\citet{2013MNRAS.430.3032B}; 
\citet{2011A&A...533A.113M}; 
\citet{2013MNRAS.428.3671T}; 
\citet{2007ApJ...656..552B}; 
\citet{2011A&A...528A..97T}; 
\citet{2016MNRAS.455.1334H}; 
\citet{2020MNRAS.491.1243P}; 
\citet{2016A&A...588L...6M}; 
\citet{2017A&A...602L..15P}; 
\citet{2017AJ....153...28S}; 
\citet{2019MNRAS.485.5168M}; 
\citet{2010ApJ...720.1118B}; 
\citet{2014A&A...563A..40C}; 
\citet{2010A&A...510A.107M}; 
\citet{2007MNRAS.375..951C}; 
\citet{2011MNRAS.413L..43P}; 
\citet{2017AJ....153..215P}; 
\citet{2014ApJ...797...42C}; 
\citet{2020Natur.580..597E}; 
\citet{2010A&A...520A..65B}; 
\citet{2019AJ....158..141Z}; 
\citet{2008ApJ...677..657J}; 
\citet{2009AN....330..475R}; 
\citet{2012MNRAS.422.3099S}
\citet{2007ApJ...671L.173B}; 
\citet{2015MNRAS.451.4060S}; 
\citet{2009IAUS..253..446H}; 
\citet{2008MNRAS.385.1576P}; 
\citet{2013AJ....145....5P}; 
\citet{10.1088/0004-637x/810/1/30}; 
\citet{2015MNRAS.454.3094S}; 
\citet{2020AJ....159..255C}; 
\citet{2015A&A...575A..61A}; 
\citet{2010PASP..122.1077D}; 
\citet{2016ApJ...820...87V}; 
\citet{2014MNRAS.437...46N}; 
\citet{2017Natur.546..514G}; 
\citet{2009Natur.460.1098H}; 
\citet{2016AJ....151..137H}; 
\citet{2021MNRAS.508.5514G}; 
\citet{2011ApJ...741..114M}; 
\citet{2019AJ....158..197R}; 
\citet{2012A&A...542A...4G}; 
\citet{2018AJ....155..119B}; 
\citet{2014MNRAS.443.2391M}; 
\citet{2019MNRAS.490.2467T}; 
\citet{2016AJ....152..182H}; 
\citet{2020A&A...643A..94L}; 
\citet{2015AJ....149..166H}; 
\citet{2019MNRAS.482.2065E}; 
\citet{2016PASP..128l4403M}; 
\citet{2019MNRAS.490.4230S}; 
\citet{10.3847/1538-3881/ab397d}; 
\citet{2020MNRAS.491.2834C}; 
\citet{2015ASPC..496..370B}; 
\citet{2012PASP..124..212S}; 
\citet{2010PASP..122.1465M}; 
\citet{2021ExA...tmp..101K}; 
\citet{2019AJ....157...82W}; 
\citet{2013A&A...554A..28C}; 
\citet{2016AJ....151...45E}; 
\citet{2016AJ....151..138R}; 
\citet{2013A&A...557A..30C}; 
\citet{2014A&A...561A..48V}; 
\citet{2013MNRAS.436....2M}; 
\citet{2014MNRAS.445.1114A}; 
\citet{2007ApJ...666L.121T}; 
\citet{2019RNAAS...3...35O}; 
\citet{2017MNRAS.468..835B}; 
\citet{2018AJ....155...79H}; 
\citet{2020AJ....159...44C}; 
\citet{2016MNRAS.459.1393M}; 
\citet{10.1088/0004-6256/141/2/63}; 
\citet{2013MNRAS.434..661C}; 
\citet{2016A&A...593A.113B}; 
\citet{2020A&A...640A..32H}; 
\citet{2019AJ....157...74A}; 
\citet{2013ApJ...764L..17B}; 
\citet{2021AJ....161....4Y}; 
\citet{2013A&A...551A..11M}; 
\citet{2016MNRAS.461.1053M}; 
\citet{2008A&A...482L..21A}; 
\citet{2018AcA....68..371M}; 
\citet{2020MNRAS.499..428S}; 
\citet{2019MNRAS.490.1294B}; 
\citet{2009ApJ...700..783P}; 
\citet{2009ApJ...693.1920H}; 
\citet{2019AcA....69..135S}; 
\citet{2021arXiv210804101C}; 
\citet{2018arXiv180904897A}; 
\citet{2010ApJ...708..224H}; 
\citet{2014AJ....147...84B}; 
\citet{2013MNRAS.430.2932M}; 
\citet{10.1088/0004-637x/773/1/64}; 
\citet{2019AJ....158...78J}; 
\citet{2016ApJ...817..141S}; 
\citet{2021RAA....21...97S}; 
\citet{2014AcA....64...27M}; 
\citet{2018PASP..130c4401C}; 
\citet{2021AJ....162....7B}; 
\citet{2013A&A...558A..86B}; 
\citet{10.1088/0004-6256/141/5/161}; 
\citet{2018MNRAS.481.4960R}; 
\citet{2019A&A...628A.116V}; 
\citet{2009AJ....137.4911F}; 
\citet{2013A&A...559A..36G}; 
\citet{2019MNRAS.490.1479H}; 
\citet{2009A&A...500L..45M}; 
\citet{2018MNRAS.480.5307T}; 
\citet{2011A&A...532A..24N}; 
\citet{2019MNRAS.485.5790T}; 
\citet{2015JATIS...1b7002S}; 
\citet{2009CoSka..39...34V}; 
\citet{2010A&A...520A..56S}; 
\citet{2010A&A...512A..14F}; 
\citet{2015MNRAS.448.2617M}; 
\citet{2011PASP..123..547M}; 
\citet{2008ApJ...686..649J}; 
\citet{2015A&A...579A.136M}; 
\citet{2010AJ....140.2007M}; 
\citet{2019MNRAS.482.1379H}; 
\citet{2011arXiv1106.4312L}; 
\citet{2018MNRAS.478.4720G}; 
\citet{2015AJ....150...33B}; 
\citet{2015ApJ...810L..23J}; 
\citet{2016PASP..128g4401C}; 
\citet{2014NewA...27..102P}; 
\citet{2017MNRAS.464..810B}; 
\citet{2014AN....335..797G}; 
\citet{2017A&A...603A..43B}; 
\citet{2015AJ....150...12B}; 
\citet{2020AJ....159..137G}; 
\citet{2017AJ....153...68S}; 
\citet{2015RAA....15..117S}; 
\citet{2016AJ....152..108E}; 
\citet{10.1093/mnras/stw574}; 
\citet{2011A&A...531A..40F}; 
\citet{2004ApJ...609.1071T}; 
\citet{2009ApJ...706..785H}; 
\citet{2015AcA....65..117S}; 
\citet{2011A&A...531A..24T}; 
\citet{2006PASP..118.1245W}; 
\citet{2008ApJ...673L..79N}; 
\citet{2018MNRAS.477.3406B}; 
\citet{2017AJ....153..211Z}; 
\citet{2007ApJ...671.2115B}; 
\citet{2007ApJ...658.1322C}; 
\citet{2015ApJ...813..111B}; 
\citet{2019PASP..131k5003A}; 
\citet{2019MNRAS.489.4125V}; 
\citet{2009A&A...507..481C}; 
\citet{2011AJ....142...86E}; 
\citet{2013PASP..125...48M}; 
\citet{10.1093/mnras/stw3005}; 
\citet{2012MNRAS.422.3099S}; 
\citet{2017NewA...55...39P}; 
\citet{2016A&A...596A..47S}; 
\citet{2017MNRAS.468.3123S}; 
\citet{2009MNRAS.392.1532J}; 
\citet{2014A&A...562L...3G}; 
\citet{2015A&A...580A..60M}; 
\citet{2004ApJ...613L.153A}; 
\citet{2019A&A...631A..76H}; 
\citet{2013MNRAS.432..944V}; 
\citet{2017A&A...600L..11C}; 
\citet{2013A&A...555A..92V}; 
\citet{2011PASJ...63..287F}; 
\citet{2010MNRAS.401.1917G}; 
\citet{2013MNRAS.431..966T}.

\bibliographystyle{aasjournal}

\end{document}